%% file: pc.tex
\documentclass{emulateapj}
\usepackage{graphicx,ifthen,url,float,color,ulem}
\def\ltsima{$\; \buildrel < \over \sim \;$}
\def\simlt{\lower.5ex\hbox{\ltsima}}
\def\gtsima{$\; \buildrel > \over \sim \;$}
\def\simgt{\lower.5ex\hbox{\gtsima}}
\newcommand {\uJy}{$\mu$Jy}
\newcommand {\um}{$\mu$m}

\newcommand {\lya}{\rm Ly$\alpha$}

\newcommand {\scuba}{{\sc Scuba}}
\newcommand {\scubaii}{{\sc Scuba-2}}
\newcommand {\planck}{{\it Planck}}

\newcommand {\spitzer}{{\it Spitzer}}
\def\um     {$\mu$m}
\def\ts     {\thinspace}
\def\kms    {\ifmmode{{\rm \ts km\ts s}^{-1}}\else{\ts km\ts s$^{-1}$}\fi}
\def\msol   {\ifmmode{{\rm M}_{\odot}}\else{M$_{\odot}$}\fi}
\def\lsol   {\ifmmode{{\rm L}_{\odot}}\else{L$_{\odot}$}\fi}
\def\zsol   {\ifmmode{{\rm Z}_{\odot}}\else{Z$_{\odot}$}\fi}
\def\etal   {{\rm et\ts al.}}
\def\aco    {\ifmmode{^{12}{\rm CO}(J\!=\!1\! \to \!0)}\else{$^{12}${\rm CO}($J$=1$\to$0)}\fi}
\def\bco    {\ifmmode{^{12}{\rm CO}(J\!=\!2\! \to \!1)}\else{$^{12}${\rm CO}($J$=2$\to$1)}\fi}
\def\cco    {\ifmmode{^{12}{\rm CO}(J\!=\!3\! \to \!2)}\else{$^{12}${\rm CO}($J$=3$\to$2)}\fi}
\def\dco    {\ifmmode{^{12}{\rm CO}(J\!=\!4\! \to \!3)}\else{$^{12}${\rm CO}($J$=4$\to$3)}\fi}
\def\gco    {\ifmmode{^{12}{\rm CO}(J\!=\!7\! \to \!6)}\else{$^{12}${\rm CO}($J$=7$\to$6)}\fi}
\def\ci     {\ifmmode{{\rm C}{\rm \small I}}\else{C\ts {\scriptsize I}}\fi}
 \def\hi     {\ifmmode{{\rm H}{\rm \small I}}\else{H\ts {\scriptsize I}}\fi}
\def\hh     {\ifmmode{{\rm H}_2}\else{H$_2$}\fi}
\def\cone {\ifmmode{{\rm C}{\rm \small I}(^3\!P_1\!\to^3\!P_0)}
     \else{C\ts {\scriptsize I}{\small$(^3\!P_1\!\to\,^3\!P_0)$}}\fi}
\def\ctwo {\ifmmode{{\rm C}{\rm \small I}(^3\!P_2\!\to\,^3\!P_1)}
     \else{C\ts {\scriptsize I}{\small$(^3\!P_2\!\to\,^3\!P_1)$}}\fi}
\def\cij {\ifmmode{{\rm C}{\rm \small I}\,(^3P_i\to^3P_j)}\else{C\ts {\scriptsize I}\,{\small$(^3P_i\to^3P_j)$}}\fi}
\def\cii    {\ifmmode{{\rm C}{\rm \small II}}\else{C\ts {\scriptsize II}}\fi}
\def\tex {\ifmmode{{T}_{\rm ex}}\else{$T_{\rm ex}$}\fi}
\def\tmb {\ifmmode{{T}_{\rm mb}}\else{$T_{\rm mb}$}\fi}
\def\tkin {\ifmmode{{T}_{\rm kin}}\else{$T_{\rm kin}$}\fi}
\def\microns {\ifmmode{\mu{\rm m}}\else{$\mu$m}\fi}
\def\nhh   {\ifmmode{n({\rm H}_2)}\else{$n$(H$_2$)}\fi}

\newcommand{\ltaraw}{$\; \buildrel < \over \sim \;$}
\newcommand{\lta}{\lower.5ex\hbox{\ltaraw}}
\newcommand{\gtaraw}{$\; \buildrel > \over \sim \;$}
\newcommand{\gta}{\lower.5ex\hbox{\gtaraw}}
\newcommand{\msun}{{\rm\,M$_\odot$}}

\newcommand{\sfr}{{\rm\,M$_\odot$\,yr$^{-1}$}}
\newcommand{\lsun}{{\rm\,L_\odot}}

\shorttitle{Assembly History of $z>2$ Proto-clusters}
\shortauthors{C.~M. Casey}
\begin{document} 

\title{The Ubiquity of Coeval Starbursts in Massive Galaxy Cluster Progenitors}
%\title{Rare galaxies as ideal beacons to the Universe's most Massive Protoclusters}
\author{Caitlin~M. Casey}
\affil{Department of Astronomy, The University of Texas at Austin, 2515 Speedway Blvd Stop C1400, Austin, TX 78712, USA}
\email{cmcasey@astro.as.utexas.edu}
%\altaffilmark{1}}
%\altaffiltext{1}{cmcasey$\@$astro.as.utexas.edu; Department of
%  Astronomy, The University of Texas at Austin, 2515 Speedway Blvd
%  Stop C1400, Austin, TX 78712, USA }

\label{firstpage}

\begin{abstract}
The Universe's largest galaxy clusters likely built the majority of
their massive $>$10$^{11}$\,\msun\ galaxies in simultaneous,
short-lived bursts of activity well before virialization.  This
conclusion is reached from emerging datasets on $z>2$ proto-clusters
and the characteristics of their member galaxies, in particular, rare
starbursts and ultraluminous active galactic nuclei (AGN).  The most
challenging observational hurdle in identifying such structures is
their very large volumes, $\sim$10$^{4}$ comoving Mpc$^{3}$ at $z>2$,
subtending areas $\sim$half a degree on the sky.  Thus the contrast
afforded by an overabundance of very rare galaxies in comparison to
the background can more easily distinguish overdense structures from
the surrounding, normal density field.  Five $2\simlt z\simlt 3$
proto-clusters from the literature are discussed in detail and are
found to contain up to 12 dusty starbursts or luminous AGN galaxies
each, a phenomenon that is unlikely to occur by chance even in
overdense environments.  These are contrasted with three
higher-redshift ($4\simlt z\simlt 5.5$) dusty star-forming galaxy
(DSFG) groups, whose evolutionary fate is less clear.  Measurements of
DSFGs' gas depletion times suggest that they are indeed short-lived on
$\sim$100\,Myr timescales, and accordingly the probability of finding
a structure containing more than 8 such systems is $\sim$0.2\%, unless
their `triggering' is correlated on very large spatial scales,
$\sim$10\,Mpc across.  The volume density of DSFG-rich proto-clusters
is found to be comparable to all $>$10$^{15}$\,\msun\ galaxy clusters
in the nearby Universe, a factor of five larger than expected in some
simulations.  Some tension yet exists between measurements of the
volume density of DSFG-rich proto-clusters and the expectation that
they are generated via short-lived episodes, as the latter suggests
only a fraction ($<\frac{1}{2}$) of all proto-clusters should be rich
with DSFGs. However, improved observations of proto-clusters
over large regions of sky will certainly shed more light on the
assembly of galaxy clusters, and whether or not they build their
galaxies through episodic bursts as suggested here.
\end{abstract}
%\begin{keywords} 
%galaxies: evolution $-$ galaxies: high-redshift $-$ galaxies: infrared $-$ galaxies: starbursts 
%\end{keywords} 

\section{Introduction}\label{introduction}

The environmental dependence of galaxies' evolution is observationally
elusive.  Locally, it is clear that galaxies residing in the most
massive environments exhibit characteristics markedly different from
their counterparts in the field: they are more massive
\citep[e.g.][]{collins09a,van-der-burg13a}, they are forming
relatively few stars \citep{balogh98a,lewis02a}, they are
preferentially red \citep{wake05a}, and they lack spiral structure
\citep{skibba09a}.  At their cores, hot inter-cluster gas$-$containing
$\sim$90\%\ of the cluster's baryonic matter$-$renders these massive
systems easy to detect via their emission of Bremsstrahlung radiation
in the X-ray \citep[see review of][]{kravtsov12a}.  These threads of
observational evidence, combined with knowledge of density
fluctuations in the early Universe imprinted on the Cosmic Microwave
Background \citep{sheth01a}, have formed the backbone of our
understanding of hierarchical growth in galaxy formation
\citep{springel05a,vogelsberger14a}.  Higher density environments saw
accelerated evolution by forming most of their galaxies early and
coalescing at earlier times. What does this imply for observations of
overdense environments at high-redshift?

In line with hierarchical expectation, some works have observed a
reversal of the star-formation-density relation at $z\sim1$
\citep{elbaz07a,cooper08b}, whereby galaxies in overdense environments
at high-redshift are {\it more} likely to be star-forming than field
galaxies or similar-mass galaxies in overdensities in the local
Universe.  However, several other works do not see this reversal
\citep{patel09a,cucciati10a,bolzonella10a,scoville13a}, leading to
some uncertainty in the processes driving evolution of clusters at
early times.

Observations of clusters in the early Universe ($z>1$) themselves also
have considerable potential as tools for testing galaxy formation
theory in a cosmological context and placing independent constraints
on fundamental cosmological parameters.  For example, discovering a
single cluster of sufficient mass at $z\ge2$ ($M_{\rm
  halo}\sim5\times10^{14}$\msun) can place significant constraints on
current cosmological models \citep[e.g.][]{harrison12a}, just as the
discovery of a population of early massive galaxies may already
challenge that paradigm \citep{steinhardt15a}.
Hence, several observational efforts to identify high-redshift
overdensities have been pursued over the past few decades
\citep{subramanian92a,steidel98a,steidel05a,ivison00b,stevens03a,miley04a,doherty10a,noble13a,rigby14a,clements14a,planck-collaboration15a}.

Unfortunately, detecting galaxy clusters at $z>2$ has proved
especially challenging.  While X-ray searches are efficient at
selecting massive clusters through emission of hot gas at
$z$\simlt$1.5$ \citep[e.g.][]{rosati02a}, the rapid surface brightness
dimming of X-ray emission makes it an inefficient observable at
high-redshift.  Other techniques for identifying cluster environments
are similarly limited to $z$\simlt$2$, such as optical searches for
the galaxy red sequence, which demonstrates the presence of an evolved
galaxy population
\citep{gladders00a,brodwin07a,eisenhardt08a,andreon14a,newman14a}, and
identifications made using the Sunyaev-Zel'dovich effect
\citep{menanteau09a,vanderlinde10a,planck-collaboration13a}.  In
addition to the difficulty in making these observations at
high-redshift, it is perhaps not surprising that these methods
struggle since they are optimized to detect {\it evolved} clusters
with older (red, massive, elliptical) galaxy populations or the
signature of a hot inter-cluster medium (ICM).  At sufficiently early
times, cluster environments may not have yet virialized to the point
where the ICM heats, implying that detection in the X-rays or 
S-Z are not optimal techniques, even if the sensitivity were
substantially deep to reach overdensities at those epochs.

Despite the difficulty in identifying clusters at high-redshift, about
twenty overdensities have been observationally identified and
spectroscopically confirmed at $z>2$.  The primary identification
technique has been targeted narrow-band filter searches around single
rare galaxies with spectroscopic redshifts \cite[e.g.][]{venemans07a}.
These narrow band imaging campaigns focus on detection of Ly$\alpha$
\citep[Lyman-$\alpha$ emitters,
  LAEs,][]{shimasaku03a,palunas04a,venemans02a,venemans05a,kuiper11a}
or H$\alpha$ \citep[H$\alpha$ emitters,
  HAEs,][]{doherty10a,hatch11a,tanaka11a,hayashi12a,koyama13a} at the
redshift of the quasar or radio galaxy.  Typically an excess of
candidate emission line galaxies is found in the vicinity of the
targeted rare source when compared against the field.  While this
constitutes strong evidence for highly clustered overdensities, most
of the emission line sources lack full spectral information or
multiwavelength characterization.

\begin{table*}
\centering
\caption{Aggregate mass, SFR, and volume characteristics of high-$z$ DSFG-rich proto-clusters}
\begin{tabular}{lccccc@{ }c@{ }c@{ }ccc}
\hline\hline
{\sc Name} & {\it z} & N$_{\rm gals}$ & $\delta_{\rm gals}$ & N$_{\rm rare}$ & $\delta_{\rm rare}$ & {\sc Ref.} & {\sc Total Stellar}  & {\sc Halo Mass}   &   {\sc Volume} & {\sc Total SFR} \\
    & &        &                 &   &     &                             & {\sc Mass} [\msun]   & (at $z$) [\msun] &   [{\it c}Mpc$^3$]            & [\sfr] \\
\hline
\multicolumn{4}{c}{Genuine DSFG-rich Proto-clusters:}\\
GOODS-N proto-cluster         & 1.99 & 34        & 2.5 & 11  & 10     & 3,\,5              & 6.5$\times$10$^{11}$     & (6$\pm$3)$\times$10$^{13}$     & 9000 & 2600$\pm$300 \\
COSMOS $z=2.10$ proto-cluster & 2.10 & $\sim$100 & 8   & 10  & 13     & 13,\,19            & 1.9$\times$10$^{12}$     & (1.7$\pm$1.2)$\times$10$^{14}$ & 15000 & 5300$\pm$600 \\
MRC\,1138$-$256 proto-cluster & 2.16 & $\sim$80  & 12  & 5   & 12     & 2,\,10,\,14        & $\sim$1$\times$10$^{12}$ & $\sim$1$\times$10$^{14}$       & 8000 & 2200$\pm$500 \\
COSMOS $z=2.47$ proto-cluster & 2.47 & 57        & 3.3 & 12  & 10     & 15,\,16,\,17       & 1.0$\times$10$^{12}$     & (8$\pm$3)$\times$10$^{13}$     & 15000 & 4500$\pm$500 \\
SSA22 proto-cluster           & 3.09 & $\sim$280 & 39  & 12  & 10     & 1,\,4,\,7,\,8,\,18 & $-$                      & (8$\pm$4)$\times$10$^{13}$     & 21000 & 5700$\pm$800 \\
\multicolumn{4}{c}{Other identified DSFG-rich Overdensities:}\\
GN20 overdensity        & 4.05 & 8         & $-$ & 3   & $>$100      & 6,\,12             & 2.8$\times$10$^{11}$     & (2$\pm$0.4)$\times$10$^{12}$   & $-^{**}$ & 1500$\pm$800 \\
HDF\,850.1 overdensity  & 5.18 & 13        & 3.6 & 2   & 6           & 11                 & $-$                      & $>$1.3$\times$10$^{11}$        & 20000 & 850$\pm$300 \\
AzTEC-3 overdensity     & 5.30 & 11        & 30  & 2   & 80          & 9                  & $\sim$2$\times$10$^{10}$ & $\sim$4$\times$10$^{11}$       & $\simgt$500 & 1600$\pm$500 \\
\hline\hline
\end{tabular}
\label{tab:allpcs}
\begin{minipage}{\textwidth}
{\small
{\bf Table Notes.} 
References are 
1=\citet{steidel98a},
2=\citet{kurk00a},
3=\citet{blain04a},
4=\citet{hayashino04a},
5=\citet{chapman09a},
6=\citet{daddi09a},
7=\citet{tamura09a},
8=\citet{lehmer09a},
9=\citet{capak11a},
10=\citet{kuiper11a},
11=\citet{walter12a},
12=\citet{hodge13b},
13=\citet{yuan14a},
14=\citet{dannerbauer14a},
15=\citet{casey15a},
16=\citet{diener15a},
17=\citet{chiang15a},
18=\citet{umehata15a}, and
19=Hung \etal, submitted.

%
%Deliniation between genuine DSFG-rich proto-clusters and strong DSFG
%overdensities which are difficult to infer collapse.
%
 $^{**}$ The GN20 structure is notably small as an association of 3
galaxies (or a total of 8, including candidates); the estimation of
its occupied volume is thus quite uncertain.  }
\end{minipage}
\end{table*}

\begin{table*}
\centering
\caption{Observed Characteristics of high-$z$ DSFG-rich proto-clusters}
\begin{tabular}{lcccc}
\hline\hline
{\sc Name} & {\it z} & {\sc Position} & {\sc Solid Angle} & {\sc Galaxy Density} \\
           &         &                & [deg$^2$]         & [deg$^{-2}$] \\
\hline
GOODS-N proto-cluster           & 1.99 & 12:36:30$+$62:13:00 & 0.17$^{o}\times$0.17$^{o}$     & 1200 \\
COSMOS $z=2.10$ proto-cluster   & 2.10 & 10:00:23$+$02:15:07 & 0.34$^{o}\times$0.13$^{o}$     & 2300 \\
MRC\,1138$-$256 proto-cluster   & 2.16 & 11:40:48$-$26:28:00 & 0.20$^{o}\times$0.10$^{o}$     & 4000 \\
COSMOS $z=2.47$ proto-cluster   & 2.47 & 10:00:31$+$02:22:22 & 0.33$^{o}\times$0.42$^{o}$     & 400 \\
SSA22 proto-cluster             & 3.09 & 22:17:34$+$00:15:01 & 0.33$^{o}\times$0.50$^{o}$     & 1700 \\
GN20 overdensity                & 4.05 & 12:37:11$+$62:22:05 & 0.01$^{o}\times$0.01$^{o}$     & 600 \\
HDF 850.1 overdensity           & 5.18 & 12:36:52$+$62:12:26 & 0.10$^{o}\times$0.13$^{o}$     & 1000 \\
AzTEC-3 overdensity             & 5.30 & 10:00:20$+$02:35:20 & 0.003$^{o}\times$0.003$^{o}$   & 1.2$\times$10$^{6}$ \\
\hline\hline
\end{tabular}
\begin{minipage}{\textwidth}
{\small {\bf Table Notes.}  Positions and observed ``sizes'' of
  high-$z$ DSFG-rich proto-clusters.  Note the large variation in
  proto-cluster solid angle and confirmed galaxy density (i.e. number
  of galaxies belonging to the proto-cluster in its solid angle).  It
  is likely that these sizes and perceived galaxy densities are
  limited by observational selection effects and are not
  representative of the structures' true physical characteristics. }
\end{minipage}
\end{table*}

In contrast, some overdensities have been serendipitously found
through large spectroscopic campaigns \citep{steidel98a,steidel05a}.
Though rare, these constitute the most spectroscopically complete
proto-clusters, some with over 100 identified LBGs or LAE member
galaxies extending several Mpc on a side.  A further handful of
proto-clusters with 5--40 LBG members have been identified surrounding
single bright submillimeter galaxies, or dusty star-forming galaxies
\citep[DSFGs, at
  $z=2-5.3$][]{chapman09a,carilli11a,capak11a,walter12a,casey15a}.

While the range of high-$z$ overdensities are diverse, this paper
focuses only on those that are spectroscopically confirmed with an
excess of DSFGs and luminous AGN.  These structures are of particular
interest as they provide unique testbeds for understanding the
assembly history of massive clusters by virtue of the presumed rarity
and short lifetimes of their constituents
\citep{solomon88a,bothwell13a,carilli13a,martini04a}.
\S~\ref{sec:observations} presents the observational characteristics
of these DSFG/AGN-rich structures.  Their potential to collapse into
some of the Universe's most massive clusters is addressed in
\S~\ref{sec:collapse}, and their unique constraints on galaxy cluster
assembly is discussed in \S~\ref{sec:simultaneous}.  Predictions are
made for the next generation of large observational surveys and
large-box simulations in \S~\ref{sec:predictions}, with conclusions
given in \S~\ref{sec:conclusions}.
Throughout, a $\Lambda$\,{\sc CDM} cosmology is assumed with $H_{\rm
  0}$=71\,km\,s$^{-1}$\,Mpc$^{-1}$ and $\Omega_{\rm m}$=0.27
\citep{hinshaw09a}, and comoving Mpc is denoted throughout as cMpc to
distinguish from proper Mpc.

\section{DSFG/AGN-rich Proto-clusters}\label{sec:observations}

Here I present the existing observational characteristics for
overdense structures at $z\simgt2$ with robust spectroscopic redshifts
and an overabundance of DSFGs or luminous AGN.  The importance of an
overabundance of DSFGs or luminous AGN is key: these are types of
galaxies that are $\simgt$100 times more rare than most `normal'
$L_\star$ galaxies across all epochs.  Their rarity is what makes them
useful for studying high-redshift overdensities, not only because they
represent a potentially critical evolutionary stage for early massive
galaxy formation \citep{toft14a}, but also because a group of them in
close proximity is exceedingly rare and can easily identify an
overdense structure too large to be identified through more common
galaxy populations.  Furthermore, as will be discussed in later
sections, they can place unique constraints on the assembly history of
proto-clusters.

Their star-formation rates, dark matter halo masses, structure
volumes, and respective population overdensities are estimated below
and discussed in context of each proto-cluster's observations.  The
star-formation rates are computed with careful treatment of individual
dust-obscured starbursts, which will dominate the calculation of SFR,
as well as a rough constraint on the contribution from other
optically-selected members like LBGs.  Dark matter halo masses are
estimated using abundance matching techniques \citep{behroozi13a},
requiring estimates to each individual member's stellar mass, unless
stated otherwise.  Due to shear numbers, the halo mass is dominated by
optically-identified member galaxies.  Third, any available
information on the physical extent of the structure is summarized,
e.g.  its occupied volume, although the uncertainty of such an
estimation should be emphasized.  Due to spectroscopic incompleteness,
all of these estimates may be viewed as lower limits in physical
terms, but can be regarded as representative of existing observable
constraints.

Galaxy overdensities are quantified with the measurement of
$\delta_{\rm gal}=(N_{\rm gal}-N_{\rm exp})/N_{\rm exp}$, where
$N_{\rm gal}$ is the observed number of member galaxies and $N_{\rm
  exp}$ is the expected number of galaxies in the same volume of blank
field, or expected cosmic density.  The expected number of galaxies is
determined using known luminosity functions for `normal' galaxies like
LBGs \citep[e.g.][]{reddy09a,van-der-Burg10a}, X-ray AGN
\citep{silverman08a} and DSFGs \citep*{casey14a}.  Different survey
depths of different fields are taken into account in determining how
prevalent a given population may be.  The observational
characteristics of all proto-clusters, as discussed in this section,
are summarized in Table~\ref{tab:allpcs}.

\subsection{GOODS-N Structure at $z=1.99$}

\citet{blain04a} and \citet{chapman09a} identified a particularly
DSFG-rich proto-cluster at $z=1.99$ in the Hubble Deep Field North.
The structure contains at least 24 optically selected,
spectroscopically confirmed members in addition to eleven rare types
of galaxies spanning the entire GOODS-N field of view (four
submillimeter galaxies, ten radio galaxies, and six X-ray galaxies,
all of which have substantial overlap).  While \citet{chapman09a}
present potentially as many as nine DSFGs, a few of those are spurious
spikes in the original SCUBA maps and others only bright radio
galaxies (Amy Barger, private communication).

\input{tabhdf}

Using deep data in HDF and more recent collections of deep
submillimeter data from {\it Herschel} \citep{oliver12a} and
\scubaii\ \citep{chen13b}, I re-derive far-infrared SEDs for this GOODS-N
structure's DSFGs using a simple modified black body and powerlaw
prescription \citep{casey12a}.  
The modified black body dominates the SED fit at rest-frame
wavelengths $\simgt$40\um, and the mid-infrared powerlaw dominates
from 5$\simgt\lambda\simgt$40\um.  Note that the calculation of
star-formation rates is largely insensitive to far-infrared SED
fitting technique, as differences in methods are typically much less
than measurement uncertainty \citep*[see \S\,4.2 of ][]{casey14a}.
The far-infrared photometry is provided
in Table~\ref{tab:hdf}.  Stellar masses and star-formation rates for
non-DSFG members are determined via detailed optical and near-infrared
SED fitting with {\sc Magphys} \citep{da-cunha08a} to rest-frame UV
data through \spitzer\ IRAC available in GOODS-N \citep{capak04a}.
The median stellar mass for non-DSFG members is
6$\times$10$^{9}$\msun\ and the median SFR is 20\,\sfr. The total
stellar mass for identified cluster members is
6.5$\times$10$^{11}$\msun\ and the total star-formation rate is
2600$\pm$300\,\sfr.  The aggregate star-formation rate is dominated
(88\%) by the DSFGs, as is the stellar mass total (70\%).

The stellar masses of the GOODS-N proto-cluster members can be checked by
extrapolating \spitzer\ IRAC photometry to rest-frame
1.6\um\ \citep{hainline09a}, as is done in \citet{chapman09a}.
Although the star-formation histories of DSFGs are quite uncertain,
and this compounds in the assumed mass-to-light ratio, I apply a
$L_{\rm H}/M_{\star}=7.9^{+0.8}_{-2.1}$\,$\lsun$/mag
\citep{hainline11a} for DSFGs and LBGs alike and derive a total
integrated stellar mass of 1.3$\times$10$^{12}$\,\msun, within a
factor of two of the SED estimate.  Treating each galaxy as its own
halo (which is appropriate given the spatial distribution of such
structures), a total dark matter halo mass is inferred for the
proto-cluster of (6$\pm$3)$\times$10$^{13}$\,\msun\ at $z=1.99$.
Assuming an exponential growth in line with large box simulations
\citep{wechsler02a}, this proto-cluster would grow to a mass of
$(9\pm5)\times10^{14}$ at $z=0$.

The comoving volume is calculated within a 10$'\times$10$'$ area and
approximate redshift bounds of $1.982<z<2.010$ as 9000\,cMpc$^{3}$.
Most of this is along the line of sight, as the spatial coverage for
deep spectra does not extend significantly beyond the deep GOODS-N
{\it HST} coverage.  %It should be emphasized for all proto-cluster
%structures, but especially for \hdfname, that the volume estimate is
%strictly a lower limit.  With wider area coverage, the structure may
%be revealed to be much larger.
It would be surprising if this structure is not extended spatially
beyond the limited field-of-view of the GOODS-N pencil-beam survey.

\subsection{COSMOS structure at $z=2.10$}

\citet{yuan14a} identify a Virgo-like progenitor in the COSMOS field
at $z=2.095$ with 57 spectroscopic members with a cluster velocity
dispersion measured to be $\sigma=552\pm52$\kms.  The proto-cluster was
revealed through spectroscopic follow-up of a zFOURGE candidate
cluster at $z=2.2$ identified with photometric techniques by
\citet{spitler12a}, and they predict a halo mass at $z\sim0$ of
10$^{14.4\pm0.3}$\msun.

Through our own Keck MOSFIRE programs to follow-up \scubaii\ and {\it
  Herschel}-selected DSFGs in the COSMOS field, there are seven
spectroscopically-identified DSFGs coincident with this structure,
four of which are published in \citet{casey12c}.  The details of this
proto-cluster, its remaining DSFGs and AGN of which there are ten
total, will be discussed in more detail in Hung \etal, submitted.
The DSFGs reach well beyond the original bounds of the structure
identified in \citet{yuan14a}, and an LBG overdensity exists across
$\sim$30$\arcmin$ scales from $z$COSMOS samples \citep{lilly09a}.  The
DSFG overdensity, centered at $z=2.10$, is measured to be $\delta_{\rm
  DSFG}=13$, with a corresponding LBG overdensity (measured from
$z$COSMOS) of $\delta_{\rm LBG}=8$.

The extensive 30$+$ bands of imaging in the COSMOS field are used to
infer stellar masses and star-formation rates from SEDs with {\sc
  Magphys}, all the details of which will be given in Hung et~al. The
aggregate stellar mass for these sources is
1.9$\times$10$^{12}$\msun\ and star-formation rate is
5300$\pm$600\,\sfr.
A lower limit on the volume for this structure is placed at
15000\,cMpc$^3$, using a sky area coverage of 8$'\times$20$'$.  While
one of the DSFGs lies significantly outside of this area, and could
easily justify a doubling of the volume, spectroscopic incompleteness
in that patch of sky significantly limits our ability to assess the
structure's extent.

\subsection{MRC1138$-$256, or the ``Spiderweb Galaxy'' structure at $z=2.16$}

This structure was originally characterized in \citet{kurk00a} and has
a number of candidate LAEs in addition to HAEs.  The most notable
member is the `Spiderweb Galaxy' described by \citet{kuiper11a}, a
radio-loud starburst with luminous AGN and giant \lya\ halo.
\citet{dannerbauer14a} present submillimeter data of the area, and
point to a number of identified DSFGs that could reside within the
structure.  From their work, five DSFGs have secure
spectroscopic confirmation within a much more spatially
compact region.  

The stellar masses of these DSFGs are estimated in
\citet{dannerbauer14a}, averaging around 10$^{11}$\msun.  Lacking
stellar mass estimates on the other spectroscopically identified
proto-cluster members, the aggregate stellar mass can be estimated 
roughly at $\sim$1$\times$10$^{12}$\msun\ and inferred halo mass of
1$\times$10$^{14}$\msun\ if abundance matching is used to separately
scale to halo mass.  This is perhaps less appropriate in this
structure than in the others given the compact spatial arrangement.  It
is possible that the mass surrounding the identified DSFGs in this
sub-halo has virialized.  Further observations will be crucial to
interpreting the size and mass of this structure (Kurk \etal, in prep)
and how it compares to the other high-$z$ structures in the
literature.

Without detailed SED information on each proto-cluster member, it is
not possible to directly derive a total star-formation rate to the
system.  However, given the far-infrared photometry provided in
\citet{dannerbauer14a}, the SFR estimates are re-derived in a
self-consistent way, and arrive at 2200$\pm$500\sfr\ as the total for
the structure.  Note that this may be overestimated due to lack of
correction for confusion boosting on the far-infrared photometry, but
may be an underestimate due to lack of inclusion of all proto-cluster
members.

The volume estimate of 3000\,cMpc$^3$ for the structure surrounding
MRC1138$-$256 uses a sky area roughly 6$'\times$9$'$ with a redshift
interval $2.154<z<2.171$.  Like the GOODS-N structure, MRC1138$-$256
is limited by a narrow field of view for multiwavelength follow-up,
and so all estimated parameters should be regarded as lower
limits, perhaps only representative of a smaller sub-halo in a larger
overdensity.

\subsection{COSMOS structure at $z=2.47$}

\citet{casey15a} describe an extended structure in the COSMOS field at
$z=2.47$ which contains seven spectroscopically-confirmed DSFGs, and
five additional AGN.  The large-field coverage of COSMOS is uniquely
useful in the identification of this overdensity, as the LBG excess is
only moderate on smaller scales ($<$1$'$).  Intriguingly, a few other
works identify a neighboring overdensity of LAEs
\citep{diener15a,chiang15a} at $z=2.44-2.45$.  While this LAE-rich
structure is offset both spatially and in redshift, by $\sim$50\,cMpc,
it could be associated as part of a colossally-large overdensity.
\citet{lee16a} detect this $z=2.44-2.45$ structure using absorption of
neutral hydrogen in the IGM, though existing data is limited to the
coincident spatial region and does not cover the $z\sim2.47$ DSFG-rich
structure.  More work is currently being carried out to determine the
possible filamentary connection between the two, and if this also
relates to a possible overdensity of DSFGs detected at $z=2.51-2.55$
in the same field.
Note that the number of galaxies in this structure has increased since
its initial publication in \citet{casey15a}; the public release of
results from the VIMOS Ultra Deep Survey (VUDS) in February 2016
revealed an additional 15 previously unidentified,
spectroscopically-confirmed proto-cluster members.

The detailed calculation of this structure's net star-formation rate
of 4500$\pm$500\sfr, total stellar mass of 1.0$\times$10$^{12}$, halo
mass of (8$\pm$3)$\times$10$^{13}$, and volume of 15000\,cMpc$^3$ is
given in \citet{casey15a} and is calculated in a fully consistent way
with the other structures described in this paper.

\subsection{SSA22 $z=3.09$ Structure}

The SSA22 structure was originally revealed in \citet{steidel98a} as
one of the first high-$z$ proto-clusters ever detected in LBGs, and as
such is probably one of the best-studied proto-clusters in the
literature. Narrow-band \lya\ follow-up has revealed an extended
excess of $z=3.1$ LAEs extending as far as 60\,Mpc comoving
\citep{hayashino04a,yamada12a,matsuda05a}.  The full extent of the
structure is shown in \citet{hayashino04a} in LAEs as reaching over
three distinct filaments about 20$'\times$3$'$, 10$'\times$4$'$ and
8$'\times$8$'$ across; the implied volume in the redshift range
$3.07<z<3.11$ is $\approx$21000\,cMpc$^3$.

\input{tabssa22}

The structure is also home to an excess of DSFGs \citep{tamura09a}.
Three DSFGs were spectroscopically confirmed as proto-cluster members
in \citet{chapman05a}, a further three were identified as
\lya\ emitters with submillimeter detections in \citet{geach05a}, and
most recently some ALMA-detected submillimeter sources have
spectroscopic confirmations from the node of the proto-cluster
\citep{kubo15a,umehata15a}.  Four of these sources are significantly
fainter than the other eight, and so are excluded from the
DSFG-overdensity calculation though are still considered for their bulk
contributions to SFR.  The FIR characteristics of these twelve DSFGs
are given in Table~\ref{tab:ssa}. Extrapolating from the 850\um\ and 1.1\,mm flux
density and a 35\,K modified blackbody template, the SFRs measured for
SSA22 DSFGs ranges from 120--1400\sfr\ and totals 5670\sfr.
In addition, there are twelve X-ray
luminous AGN present in the proto-cluster \citep{lehmer09a}, four of
which overlap with the DSFGs, bringing the total rare galaxy count to
13. \citet{lehmer09a} also finds evidence that the LBGs in SSA22 are a
bit more massive (by factors of 1.2--1.8) than LBGs in the field, and
\citet{hine16a} shows evidence for enhanced merger rates in
proto-cluster member galaxies.

It is difficult to precisely identify how many spectroscopically
confirmed proto-cluster members sit in the SSA22 proto-cluster.  The
original spectroscopic sample has only 16 members, while the
narrow-band follow-up imaging around Ly$\alpha$ has 283 confident
candidates extending $\sim$half a degree across the sky. In addition,
there have been several further spectroscopic campaigns in the field,
confirming a handful of interesting sources.  No stellar mass
estimates are given for this structure, although \citet{steidel98a} do
provide an estimate of the total halo mass of
(8$\pm$4)$\times$10$^{13}$\msun\ computed using the implied bias from
the LBG overdensity.

\subsection{The GN20 overdensity at $z=4.05$}

One of the brightest submillimeter galaxies from the original
\scuba\ surveys, GN20 eluded redshift identification for many years
until \citet{daddi09a} confirmed it at $z=4.055$ through a
serendipitous CO detection.  Follow-up work revealed two accompanying
galaxies, themselves submillimeter emitters, at the same redshift.
This GN20 system is discussed in detail in \citet{hodge13b}.  This
overdensity is significantly different than the structures discussed
so far, with far fewer proto-cluster members identified through
spectroscopy.  This may indicate that it is intrinsically less massive
than the other structures, or that spectroscopic incompleteness is
quite severe.  Because the structure sits in the well studied GOODS-N
field (like the GOODS-N structure at $z=1.99$) spectroscopic
incompleteness is less likely, particularly at a redshift where
detecting Ly$\alpha$ emitters would be fairly straightforward with
ground-based optical multi-object spectrographs
\citep{wirth04a,cowie04a}. 

Stellar mass estimates for this group are given in \citet{daddi09a}
and \citet{hodge13b} for the three DSFGs: GN20, GN20.2a, and GN20.2b.
The sum of their stellar masses is $\sim$3$\times$10$^{11}$\msun, and
total star-formation rate of 1500$\pm$800\sfr.  Hodge \etal\ reveal
six tentative CO(2-1) detections surrounding the GN20 complex, and the
50\,cMpc$^{3}$ volume for the structure is thus estimated within a
4$'\times$3$'$ area and a redshift interval of $\Delta z=0.0014$ at
$z=4.055$.  Like the lack of large numbers of spectroscopic
confirmations, the estimated volume is quite a bit smaller than the
other structures presented here, which may be due to the fact that we
are looking at a sub-halo in a larger structure, or more likely,
a group which is intrinsically less massive than the five
structures presented so far that sit at lower redshift.

\subsection{The HDF\,850.1 overdensity at $z=5.18$}

\citet{walter12a} describes the massive starbursting submillimeter
galaxy HDF\,850.1 and the structure surrounding it at $z\approx5.2$.
Like GN20, HDF\,850.1 eluded redshift confirmation for over a decade
and was only confirmed via detection of molecular gas.  While it is
the only DSFG in this $z=5.2$ overdensity, there is an accompanying
QSO and eleven other spectrosccopically-confirmed galaxies at the same
redshift.  This overdensity extends across a large filamentary area
10$'\times$30$'$.  Its total star formation rate is estimated just
using the single submillimeter source for lack of adequate photometric
constraints on the other proto-cluster members, at 850$\pm$300\,\sfr.
Similarly, given the high redshift of this structure, stellar masses
are unconstrained due to lack of atmospheric transmission around
rest-frame 1.6\um. Do note, however, that there is a dynamical
mass constraint on the galaxy HDF\,850.1 of
(1.3$\pm$0.4)$\times$10$^{11}$\msun, which can be used as a lower limit
to the halo mass of the system at $z\approx5.2$.  The volume estimate
of 20000\,cMpc$^{3}$ is derived assuming the above solid angle and a
redshift range of $5.183<z<5.213$.

\subsection{The AzTEC-3 overdensity at $z=5.30$}

\citet{capak11a} report the discovery of an overdensity surrounding
the interesting luminous DSFG named AzTEC-3 in the COSMOS field.
Within a 1$'$ diameter region, there appear to be twelve proto-cluster
members at $z\approx5.3$, including the single DSFG AzTEC-3 and one
X-ray detected quasar at a distance of 13\,Mpc from the starburst.
Similar to the HDF\,850.1 overdensity, estimating stellar masses for
these sources is quite challenging, although \citet{capak11a} offer
this computation directly, totaling
$>$2$\times$10$^{10}$\msun.  They extrapolate this to a halo mass
using abundance matching techniques and estimate a lower limit of
$>$4$\times$10$^{11}$\msun.  The total SFR estimate is again taken for
the sole DSFG member at 1600$\pm$500\sfr.  The volume of
the structure is estimated within a 0.5$'$ radius and a $\Delta z=0.03$ interval,
arriving at a lower limit of $\simgt$500\,cMpc$^3$.
As is the case with the other high-redshift overdensities, it is
important to stress that the AzTEC-3 system could be the progenitor of
a less massive overdensity.

\subsection{Candidate DSFG-rich Proto-clusters}

It is important to emphasize again that a number of candidate
high-$z$, DSFG-rich proto-clusters have recently been found thanks to
wide-area surveys like those from {\it Planck} and {\it Herschel} but
are awaiting spectroscopic confirmation
\citep{clements14a,planck-collaboration15a,flores-cacho16a}.  It is
similarly important to stress that not all other
spectroscopically-identified $z>2$ proto-clusters have the sensitive
submillimeter datasets needed to detect potential DSFG member
galaxies \citep[e.g.][]{lee14a}. 

\section{From DSFG-rich Proto-clusters to $z\sim0$ Clusters}\label{sec:collapse}

Such physically large, extended structures $-$ like those
observationally identified in \S~\ref{sec:observations} $-$ are not
certain to collapse into massive galaxy clusters.  How can we
adequately determine whether or not these structures will collapse by
$z\sim0$?  And does their number density agree with what is known
about galaxy clusters at $z\sim0$?

\subsection{Will they collapse?}

Two schools of thought have been used to address this question.  The
first draws on the Press-Schechter formalism \citep*{press74a} for
spherical collapse within large scale structure \citep*{mo96a},
whereby a certain mass overdensity, $\delta_{\rm mass}$, is required
to exceed a specific critical value $\delta_{\rm c}$ to collapse by
$z\sim0$ \citep{peacock99a}.  Because the mass overdensity is not
directly observable, linear galaxy bias is assumed whereby
$1+b\delta_{\rm mass}=C(1+\delta_{\rm gal})$, and $\delta_{\rm gal}$
is the observed galaxy overdensity, $b$ is the bias associated with
that galaxy population (i.e. how well they trace the dark matter halo
mass), and $C$ is a redshift distortion factor accounting for unknown
peculiar velocities.

For example, the analysis of the GOODS-N $z=1.99$ structure in
\citet{chapman09a} finds an SMG overdensity of $\delta=10$, sufficient
to cause collapse, however the underlying LBG population overdensity,
$\delta_{\rm LBG}=2.5$, is not significant enough to cause collapse.
These two assessments of the structure are seemingly contradictory,
but the authors address this contradiction by suggesting that either
the bias of the submillimeter galaxy population is sufficiently
different than for LBGs, or there could be a large population of
massive galaxies that have not been detected surrounding the
structure.  Given the depth of multiwavelength imaging in GOODS-N the
latter is unlikely.  Thus \citeauthor{chapman09a} determined that the
bias for SMGs (or DSFGs) and LBGs was sufficiently different, and so
even a large overdensity of SMGs may not probe massive clusters in
formation.

\begin{figure}
\includegraphics[width=0.99\columnwidth]{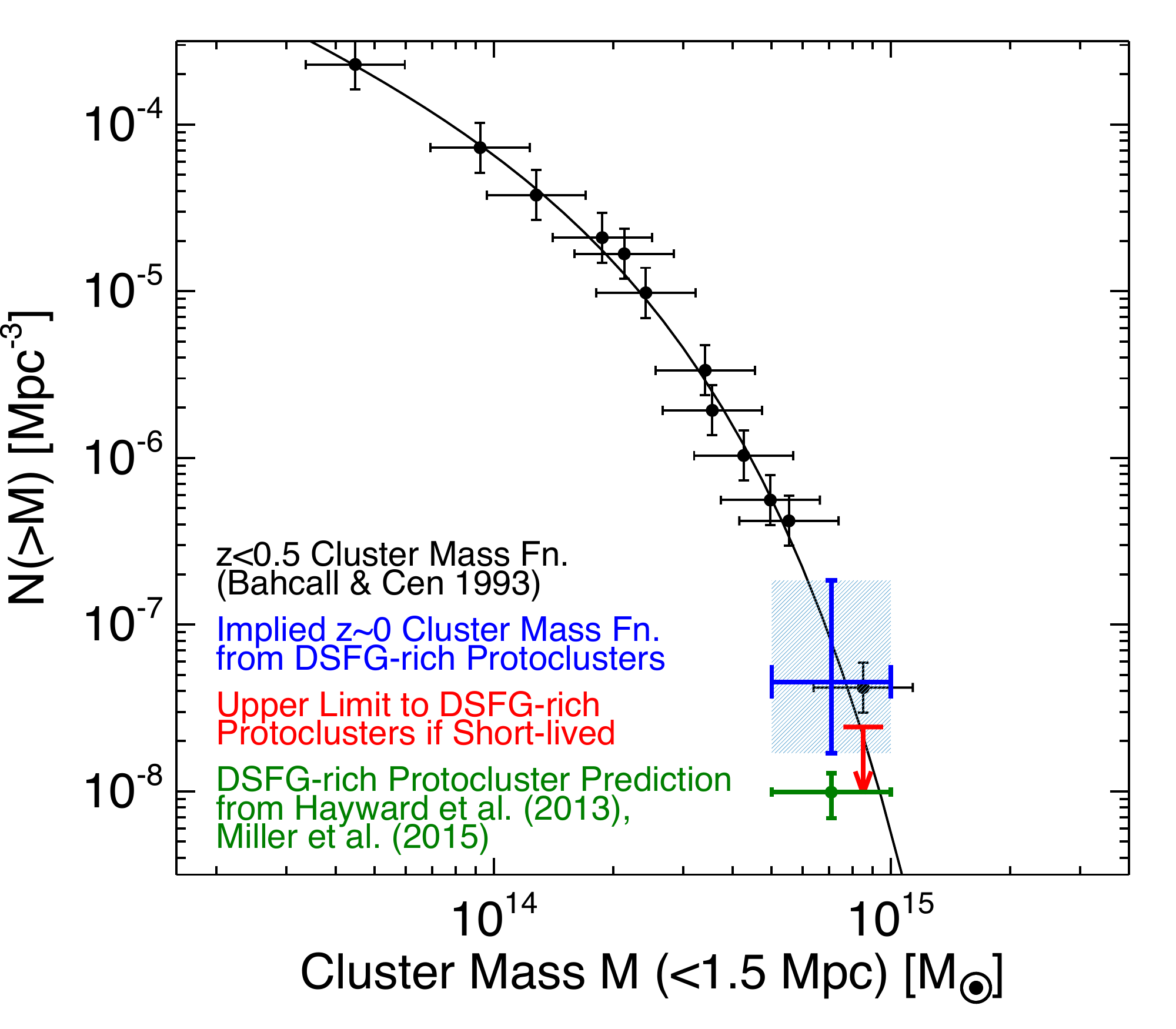}
\caption{A comparison of the Cluster Mass Function from
  \citet{bahcall93a} against the extrapolated estimate of DSFG-rich
  proto-cluster descendants (blue point).  The uncertainty is dominated
  by the limited understanding of the DSFG-rich proto-cluster selection
  function, and the survey area from which they have been found;
  future large-area surveys will enable a much more accurate
  constraint.  If DSFG-rich proto-clusters are assumed to be a
  short-lived phenomenon, then an upper limit (red arrow) marks
  maximum density for DSFG-rich structures.  The estimate of DSFG-rich
  proto-clusters (i.e. those with five or more DSFGs) from large-box
  {\it Bolshoi} simulations is shown as the green point
  \citep{hayward13a,miller15a}. }
\label{fig:cmf}
\end{figure}

This conclusion is further supported in \citet{miller15a} who use
large-volume semi-analytic simulations from \citet{klypin11a} to argue
that SMGs are ``poor tracers'' of the most massive structures at
$z\sim2$, observing very few massive structures containing more than
1--3 SMGs.  The structures observed with $>$5 SMGs are indeed amongst
the most massive, but are exceedingly rare in the simulation, much
more so than the observations in \S~\ref{sec:observations} suggest.
This discrepancy between their predicted number of DSFG-rich
proto-clusters and our observations are shown as green and blue points
on the cluster mass function plot in Figure~\ref{fig:cmf}, discussed
in more detail in the next subsection.

Note that other simulations groups \citep{granato15a,lacey15a} have
been working to understand the turn-on of luminous DSFGs in large-box
simulations where the collapse of the most massive structures can be
seen.  The advantage of these techniques is the ability to directly
constrain SMGs' physical drivers, which they largely attribute to disk
instabilities and a mildly top-heavy IMF.  However, as highlighted in
those works, it is still incredibly challenging to carry through
proper radiative transfer in such large environments, especially on
$\sim$20\,cMpc scales before clusters have collapsed.

The second school of thought draws on recent cosmological simulations
of hierarchical growth, which produce somewhat different predictions
than those relying on analytic descriptions of structure formation
theory.  For example, \citet{chiang13a} present a clear argument as to
why spherical collapse models and the assumed linear regime for
overdensities may introduce systematic errors in mass measurements for
non-virialized proto-clusters.  These direct predictions from
simulations suggest that: (a) the median observed galaxy overdensity,
$\delta_{\rm gal}$, rarely, if ever, exceeds 10 \citep[this agrees
  with the predictions given in][]{miller15a}, (b) $\delta_{\rm gal}$
at these epochs also depends strongly on the observational
characteristics being selected for, for example SFR or stellar mass,
and sensibly vary between DSFG populations (very high SFR-selected
samples) and LBG populations (a combination of SFR and mass selected,
at much deeper detection thresholds), (c) the progenitors of massive
galaxy clusters at $z>2$ occupy very large Lagrange volumes,
$\simgt$10000\,cMpc$^3$ \citep[see also][]{onorbe14a}, and (d)
$\delta_{\rm gal}$ will vary for structures of the same mass depending
on the `window size' of observations, or presumed volume, given
intrinsic variations in the underlying density along filaments.

For example, a close inspection of Figure 8 in \citet{chiang13a}$-$ a
plot of the cumulative fraction of proto-clusters with observed galaxy
overdensities $\delta_{\rm g}$ at $z=2$, 3, 4 and 5$-$provides a
backdrop to interpret the likelihood of proto-cluster collapse.  Among
the five rich $1.99<z<3.09$ proto-clusters described in
\S~\ref{sec:observations}, all structures are expected to collapse by
$z\sim0$.  The structure with the least remarkable LBG overdensity at
$\delta_{\rm gal}=2.5$, the GOODS-N $z=1.99$ structure, is still among the
top 30\%\ of collapsing structures.  The remainder are in the top
5--10\%.

It should be clarified, however that the three highest redshift
overdensities discussed in \S~\ref{sec:observations} and summarized in
Table~\ref{tab:allpcs} have a less clear fate.  With far fewer numbers
of galaxies (in both rare sub-types and total number), Poisson noise
dominates the calculation of the overdensities, causing a wide margin
of error on the structures' predicted state at $z\sim0$.  These are
the types of structures which may either be prone to mass
overestimation, due to the effects discussed in \citet{miller15a}, or
suffer from incomplete spectroscopic descriptions, though the latter
interpretation may be limited by constraints set by the volume density
of DSFG-rich proto-clusters as a whole.

\subsection{How common are they?}\label{sec:common}

While the argument for the eventual collapse of DSFG-rich
proto-clusters into the most massive $z\sim0$ clusters has been made in
\S~\ref{sec:collapse}, it is not immediately obvious that this
evolutionary picture is feasible or likely, given the relatively small
number of high-mass clusters at $z\sim0$.  In Figure~\ref{fig:cmf},
the cluster mass function at $z\simlt0.2$ is shown from the Sloan
Digital Sky Survey \citep{bahcall93a,bahcall03a}.  This tells us that
there is one $>$10$^{15}$\,\msun\ cluster per every 1--2 million
Mpc$^3$, or per $\sim$120$\times$120$\times$120\,Mpc comoving box.

We can also work out a rough estimate to the volume density of
DSFG-rich proto-clusters for comparison.  A significant discrepancy
between the volume density of $z\sim0$ massive clusters and $z\simgt2$
DSFG-rich proto-clusters is a sign that the two populations are not
likely related\footnote{Either they are not likely related, or if they
  are DSFG-rich proto-clusters are probably much more rare than most
  `normal' proto-clusters.}.  This was claimed to be the case in
\citet{blain04a}, after analyzing the overdensity associated with the
GOODS-N $z=1.99$ structure, and a few other potential SMG-rich
overdensities perceived in the first few square degrees of deep submm
imaging.  \citeauthor{blain04a} determined that DSFG-rich structures
were unlikely to be the progenitors of massive clusters in formation
because they are $\sim$10 times more common at $z\simgt2$ than their
$z\sim0$ descendants, which was reflective of the best data on-hand at
the time.  Here this estimation is reassessed with improved datasets.

\begin{figure*}
\centering
\includegraphics[width=1.7\columnwidth]{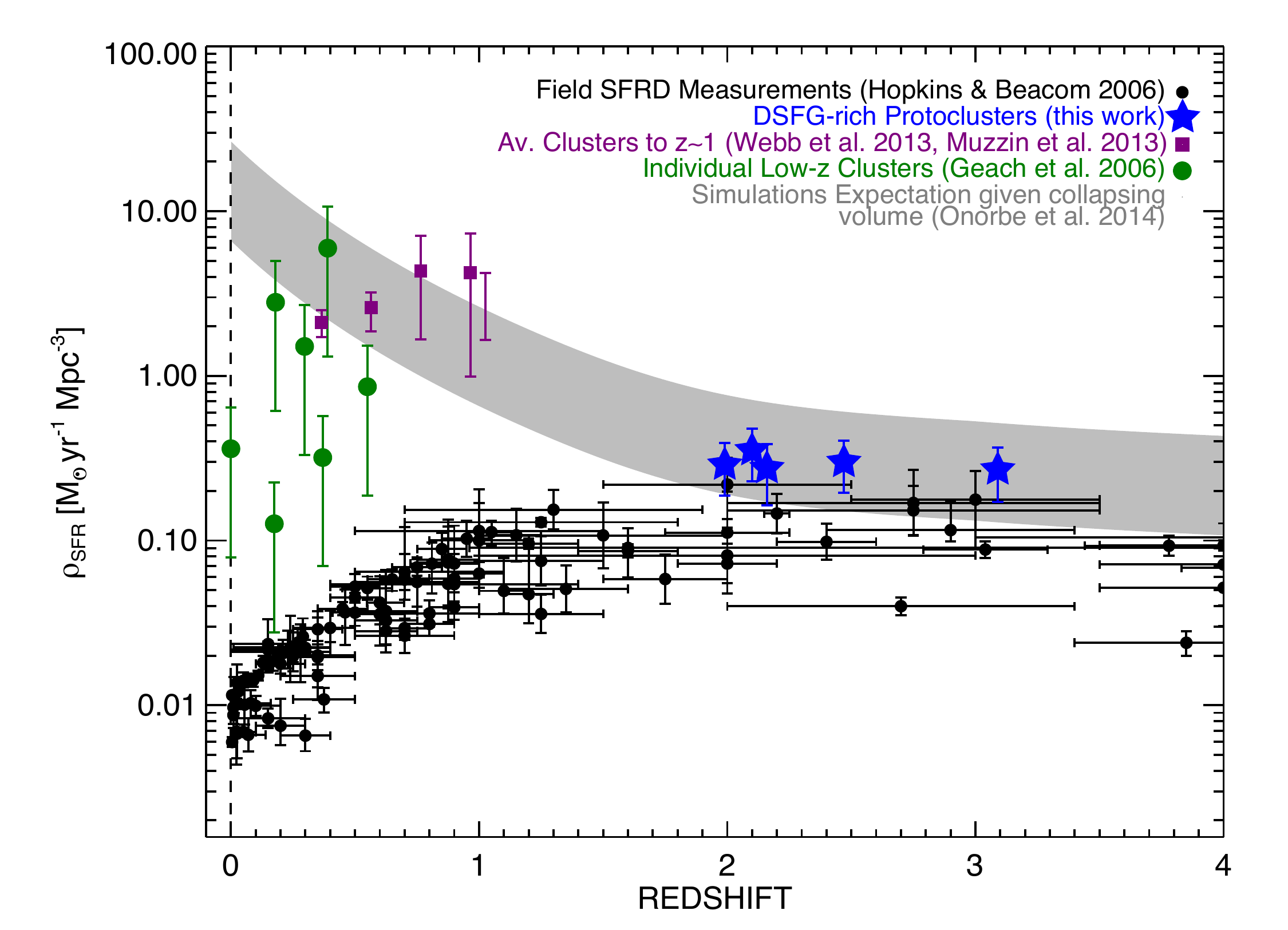}
\caption{The star-formation rate density ($\rho_{\rm SFR}$, in
  \sfr\,Mpc$^{-3}$) of proto-clusters and clusters, in comparison to
  galaxies in the field from \citet{hopkins06a} (black points).  The
  five DSFG-rich proto-clusters from this work are shown as blue stars,
  seven individual low-redshift clusters from \citet{geach06a} are
  shown as green circles, and redshift-averaged results from 42
  clusters at $0.3<z<1.0$ in \citet{webb13a} shown in purple.  The
  virialized clusters have an $\sim$2\,Mpc proper radius, and the
  associated volume is converted into comoving units for fair
  comparison with the field and proto-clusters.  The gray stripe
  represents the track of a hypothetical and idealized proto-cluster
  which sustains a constant SFR$\approx$3000$\pm$1500\sfr\ from
  $z\sim4$ to $z=0$, but whose SFR density increases by a factor of
  $\sim$100 from $z=2$ to $z=0$ due to the dramatic reduction in
  comoving volume as the cluster collapses and virializes.  Real
  clusters may see this steep rise in $\rho_{\rm SFR}$ from $z=2$ to
  $z=1$, but then experience some type of quenching which cuts off
  ongoing star-formation at more recent times $z<1$, as demonstrated
  by some of the low-$z$ clusters shown in green and purple here.
}
\label{fig:sfrd}
\end{figure*}

To estimate the volume density of DSFG-rich proto-clusters at
$z\simgt2$, understanding survey area and selection bias is critical.
Survey area in this case is set by the solid angle of sky covered to
sufficient depth to recover DSFG-rich structures at high-redshift.
This requires both spectroscopically complete samples {\it and}
confusion-limited submillimeter blank-field maps.  Both are extremely
limited by current observational resources. The former is limited by
the need for several tens of nights on 8--10\,m class
optical/near-infrared telescopes for multi-object spectroscopy of
faint $i\sim22-26$ sources (of which only a few fields have truly
complete coverage, e.g. GOODS-N, central portion of COSMOS, ECDF-S),
and the later is limited by the historically slow mapping speeds of
single-dish bolometer array instruments like SCUBA (also LABOCA,
MAMBO, AzTEC, and now SCUBA-2).  The intersection of these two
datasets is therefore limited to: 
\begin{itemize}
\item about 0.4\,deg$^{2}$ in GOODS-N,
a field with confusion limited 850\um\ data \citep{barger98a,chen13b}
and extensive spectroscopic completeness
\citep{cowie04a,wirth04a,reddy06a,barger08a},
\item the central 1\,deg$^{2}$ of the COSMOS field, which has
  published confusion-limited 850\um\ data covering 0.2\,deg$^2$ (more
  yet in Geach \etal, in prep), and 1\,deg$^2$ of deep spectroscopic
  data from the $z$COSMOS team \citep{lilly09a},
\item about 0.5\,deg$^2$ in ECDF-S with
confusion-limited submm data from LABOCA and ALMA
\citep{weis09a,hodge13a} and spectroscopic follow-up from
\citet{popesso09a}, \citet{balestra10a}, and \citet{le-fevre05a}, and
\item a 0.5\,deg$^2$ portion of the Lockman Hole (SHADES) field with for
which a significant number of DSFGs have been spectroscopically
confirmed \citep{chapman05a,lindner11a,casey12b}, and
\item other deep submillimeter fields, which include the backgrounds
  of low-redshift Abell clusters \citep[e.g.][]{smail97a,chen13a} and
  the SSA13 and SSA22 fields, and cumulatively add up to about
  $\sim$0.5\,deg$^2$.
\end{itemize}
This collection of deep surveys adds up to a total effective survey
solid angle of $\sim$3\,deg$^2$, with an uncertainty of about
$\sim$0.5\,deg$^2$ to account for variable levels of spectroscopic
completeness and submm data quality and depth.  
While it should be noted that {\it Herschel} coverage also spans all
of these legacy fields, the intersection with spectroscopic samples is
the main limiting factor in making use of it for this analysis.  In
addition, {\it Herschel} is most efficient at identifying DSFGs at
$z<2$ \citep{casey12b}, a characteristic of its shorter-wavelength
selection than ground-based submm datasets.  Color selection with the
{\it Herschel} bands seems like an efficient method of recovering a
higher-redshift sample \citep[e.g.][]{dowell14a,asboth16a}, though the
depth and completeness of these `500\um-peakers' is less well
characterized.
It is important to emphasize that this estimation is very rough, as
the complexity of these datasets is incredibly difficult to quantify
in a simple analysis.

The corresponding solid angle to this 3\,deg$^2$ is converted to a
cosmological comoving volume within the redshift interval of interest,
which is approximated as $1.9<z<4.5$, the lower limit defined by the
limit of known virialized clusters and the upper limit constrained by
low completeness in most large spectroscopic surveys summarized above.
Allowing for some additional uncertainty in the redshift interval, the
total volume accessible is (9$\pm$3)$\times$10$^{7}$\,cMpc$^{3}$.  By
chance this is approximately the same volume probed by deep SDSS
cluster surveys, $\sim$400\,deg$^{2}$ out to $z\sim0.1-0.2$
\citep{bahcall03a}, from which the nearby cluster mass function is
measured.

Though there are clearly these five, bona-fide DSFG-rich proto-clusters
identified in the literature, one is substantially impacted by a
possible selection bias associated with the proto-cluster.  Much of the
deep data associated with MRC1138$-$256 at $z=2.16$ was obtained with
the explicit knowledge of the proto-clusters' presence, and so it
cannot be included in the calculation estimating their volume density.
Thus four DSFG-rich proto-clusters are left for the volume density
calculation: GOODS-N at $z=1.99$, COSMOS at $z=2.10$, COSMOS at $z=2.47$,
and SSA22 at $z=3.09$. A Poisson uncertainty is assumed for the number
of DSFG-rich proto-clusters.  The implied volume density is then
$\sim$5$\times$10$^{-8}$\,cMpc$^{-3}$ for DSFG-rich proto-clusters.
This is depicted by the blue point on Figure~\ref{fig:cmf} and is in
rough agreement with the observed $z\sim0$ cluster mass function.
%contrary to the earlier results of
%\citet{blain04a}.

There is one remaining concern with this calculation.  If this
estimate is consistent with the $z\sim0$ local cluster mass function,
then it may imply {\it every} $z\simgt2$ proto-cluster should be
DSFG-rich.  This is not obviously the case.  Before addressing this
issue further, one must first consider the timescale of DSFGs and their
implications on clusters' assembly histories.

\subsection{Star-Formation in DSFG-rich Proto-clusters}

Placing DSFG-rich proto-clusters in context requires a more detailed
look at their observable star-formation characteristics in comparison
to the field (i.e. normal density regions), and lower redshift
virialized clusters. Figure~\ref{fig:sfrd} shows the cosmic
star-formation rate density from $0<z<4$ as compiled by
\citet{hopkins06a} for the field, against similar measures for
overdense environments.  

DSFG-rich proto-clusters at $2<z<3$ only have slightly elevated
$\rho_{\rm SFR}$ than the field, thanks primarily to the large volumes
they occupy prior to virialization.  On the other hand, virialized
clusters at $z<1$ have substantially higher $\rho_{\rm SFR}$, peaking
around $0.5<z<1.0$, while potentially experiencing suppressed
star-formation at lower redshifts brought on by different
environmental mechanisms.  Note that the comparison between virialized
clusters and the field uses comoving volume, as opposed to proper
volume, for fair comparison with structures which have not yet
collapsed and decoupled from the Hubble flow.  All values of SFR are
converted to a Chabrier IMF \citep{chabrier03a}.  The gray band marks
the evolution of a hypothetical cluster that sustains an aggregate SFR
of 3000\sfr\ from $z=4$ to $z=0$ while undergoing collapse as
predicted from large N-body simulations \citep{onorbe14a}.  This
highlights, through one variable, how galaxies in proto-clusters more
closely emulate galaxies in the field than those in $z\sim1$ clusters
that have collapsed.

\begin{figure}
\includegraphics[width=0.99\columnwidth]{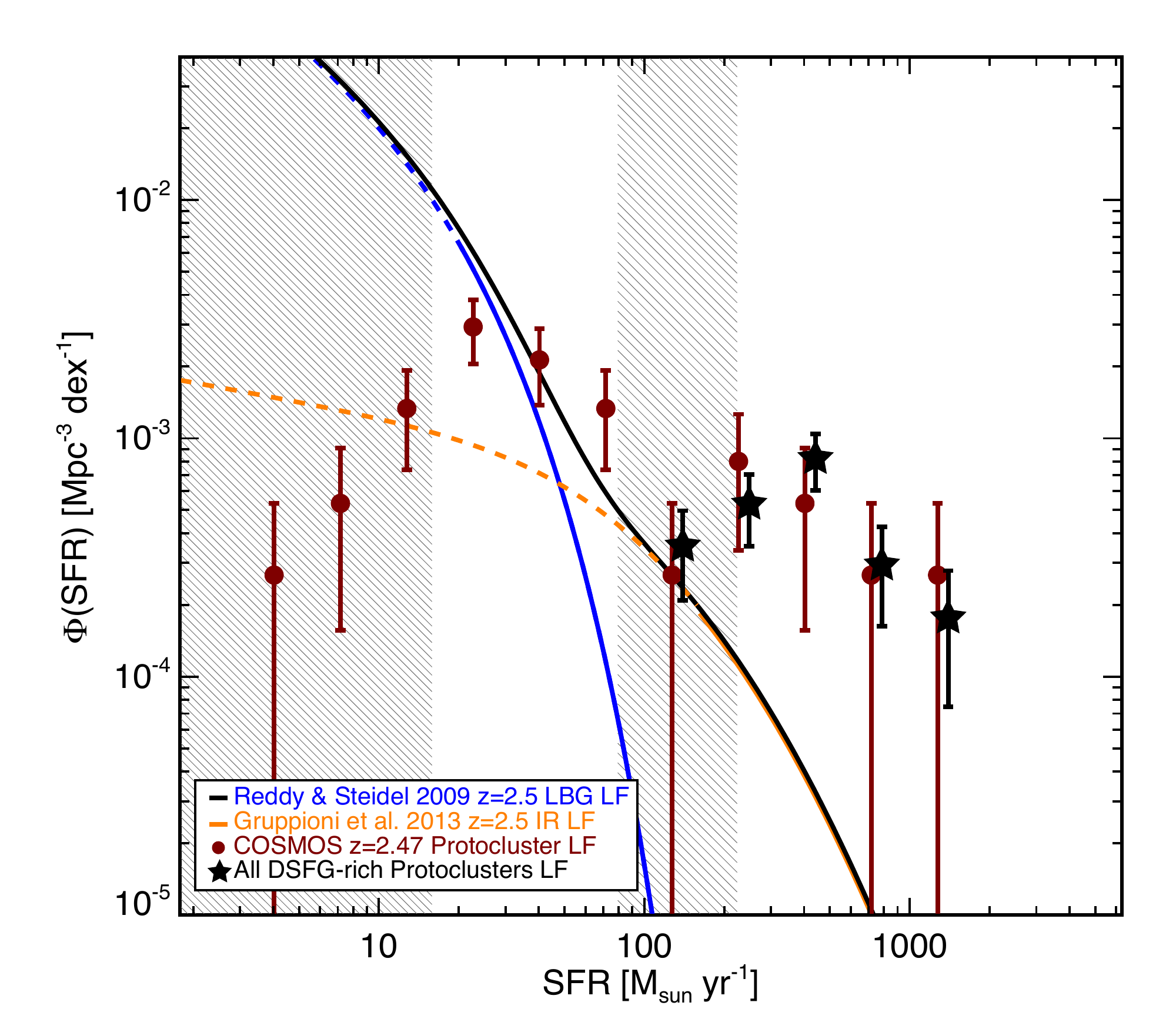}
\caption{The star-formation rate function of DSFG-rich proto-clusters
  compared to the field.  The luminosity functions of Lyman-break
  galaxies \citep[blue line;][]{reddy09a} and IR-selected galaxies
  \citep[orange line;][]{gruppioni13a} in the field are shown for
  context; the black line sums the two.  The SFR function of the
  COSMOS $z=2.47$ proto-cluster (red points) is shown for all known
  members. No correction has been made for incompleteness (hashed gray
  regions and dashed lines), which dominates at
  SFRs$\simlt$20\,\sfr\ (for UV-selected samples) and at
  SFR$\approx$100-200\,\sfr\ (for DSFGs). The net SFR function for all
  five DSFG-rich proto-clusters at $1.99<z<3.09$ is shown as black
  stars.}
\label{fig:sfrlumfn}
\end{figure}

Figure~\ref{fig:sfrlumfn} takes a closer look at the breakdown of the
star-formation rate function, or luminosity function within a
DSFG-rich proto-cluster in comparison to the field.  For context, the
Lyman-Break Galaxy luminosity function of \citet{reddy09a} is
converted to a SFR function using the UV-scaling in
\citet{kennicutt98b} and applying a factor of five correction for
extinction \citep[i.e. most LBGs are 80\%\ obscured;][]{reddy12a}.
The highest redshift luminosity function from the infrared
\citep{gruppioni13a} is converted to a SFR also using the Kennicutt
prescription, adjusted for a Chabrier IMF.  Against these field
measurements, the SFR function of DSFG-rich proto-clusters is shown:
all of the known members of the COSMOS $z=2.47$ proto-cluster
\citep{casey15a} in red, and the DSFG member galaxies of all five
$1.99<z<3.09$ structures in black stars.  The key distinguishing
characteristic of DSFG-rich proto-clusters is the flattening of the
luminosity function towards high SFRs.
%, such that
%the excess of galaxies above expectation is more strongly pronounced.
While there may be an excess of LBGs observed in high-$z$
proto-clusters, the excess is not as great as the factor $\simgt$10
excess towards the highest SFRs.

\section{Simultaneous Triggering, or not?}\label{sec:simultaneous}

Here the likelihood of several rare types of galaxies being observed
simultaneously within a large structure is explored.  If you work from
the premise that both populations of DSFGs and AGN are short-lived on
100\,Myr timescales, then one can ask what the probability is of
observing $N$ of them simultaneously in one structure (where
$N\simgt5$).  If the probability is low, and yet the prevalence of
such DSFG-rich structures is high, then one may think this is evidence
that clusters themselves assemble in rapid bursts, even when extended
over very large volumes $\simgt$10000\,$c$Mpc$^3$ \citep[as suggested
  in][]{casey15a}.

Care should be taken in correcting for the dynamical time of each
DSFG at different redshifts, as discussed in \citet{simpson14a}.
 At higher redshifts, a fixed $dz$ element probes shorter and
shorter timescales, such that the probability of observing {\it all}
DSFGs which have been triggered during that time element $dz$
increases from low fractions at low-$z$ to 100\%\ at high-$z$.  While
large redshift bins with widths $\Delta z=0.1-0.2$ will probe all such
episodes, it is important to note that the redshift range probed by a
single coherent structure, $dz\approx0.02$, only corresponds to a
crossing time of $\approx\,20$\,Myr, shorter than the expected duration
of the burst phase.  If this itself were to exceed the estimated
lifetimes of our rare galaxies, that could provide an easy explanation
as to why we observe structures that are quite rich in DSFGs and
luminous AGN.  However, that is not the case.

Another possible explanation for the plethora of rare galaxies is that
we actually expect nearly all $z\sim0$ galaxy cluster members to have
gone through such a rare phase at some time in its past, probably
around $z\sim2-3$.  But in investigating this further, there is a
problem.  The most massive galaxy clusters at $z\sim1$ only have
40$\pm$10 galaxies above a stellar mass of
10$^{11}$\msun\ \citep{van-der-burg13a}.  If one presumes all of these
have gone through a DSFG phase at some point during their mass buildup
(as most of them are quiescent by $z=0.5-1$), then by working
backwards, the likelihood of observing $N$ of them in a DSFG or
luminous AGN phase simultaneously can be worked out.  Here the time
$T$ it takes for the structure to collapse from its primordial
fluctuations is relatively unknown, but is loosely constrained by the
redshift interval $1<z<6$ ($\approx$5\,Gyr), or $2<z<5$
($\approx$2\,Gyr).

\begin{figure}
\includegraphics[width=0.99\columnwidth]{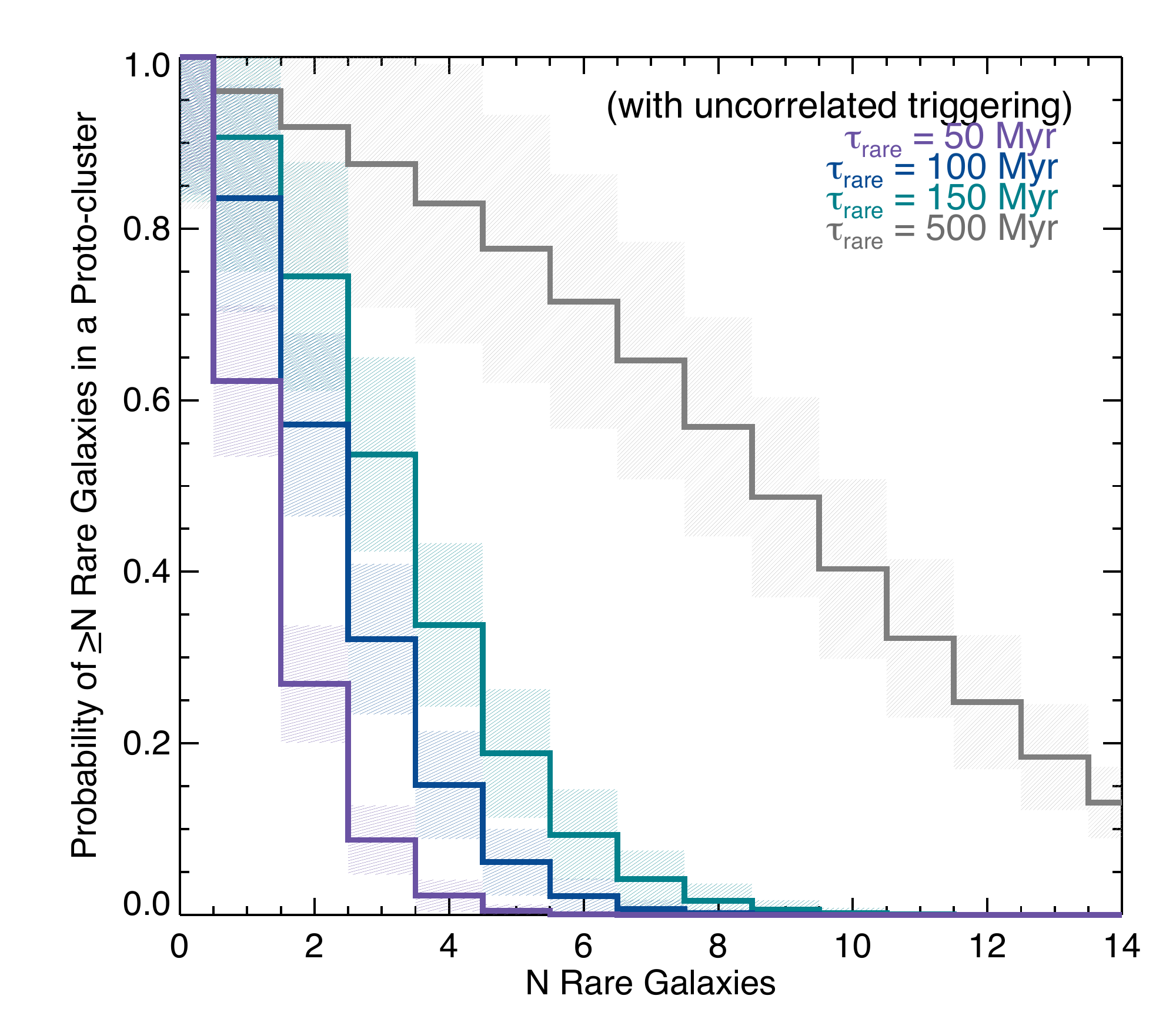}
\caption{The probability of observing $\ge N$ rare galaxies (including
  DSFGs and luminous AGN) simultaneously in one proto-cluster
  structure, if uncorrelated, random triggering is assumed.  With
  40$\pm$10 massive $>$10$^{11}$\,\msun\ galaxies in present-day
  massive galaxy clusters \citep{van-der-burg13a}, one can assume all
  of them passed through a DSFG phase at some point between $1<z<6$ in
  which they build the majority of their mass. If the rare galaxies
  are triggered at random during that time/redshift interval
  (i.e. they are uncorrelated events) the likelihood that $N$ or more
  of them would be `on' simultaneously is shown, given an average
  lifetime of 50\,Myr (purple), 100\,Myr (blue), 150\,Myr (teal), or
  500\,Myr (gray).  For example, if DSFGs are short-lived, the
  likelihood of observing $>$3 per structure is very low.  Conversely,
  if DSFGs are long-lived, we are more likely to see structures with
  $\ge$8 DSFGs than with fewer.}

\label{fig:probpern}
\end{figure}

Figure~\ref{fig:probpern} shows the probability of simultaneously
observing $\ge N$ DSFGs/AGN within one structure forming over the
course of 2\,Gyr.  Assuming a 2\,Gyr timescale renders the probability
calculations in Figure~\ref{fig:probpern} conservative, as allowing
for longer fall-in times makes the probabilities of observing multiple
DSFGs simultaneously only lower.
Four different rare-galaxy timescales are assumed (where ``rare'' can
refer to either the DSFGs or the short-lived, luminous AGN in this
case): 50\,Myr \citep[in line with what is observed in local
  ULIRGs;][]{solomon88a}, 100\,Myr \citep[the typical DSFG timescale
  and upper limit to QSO lifetimes;][]{greve05a,martini04a}, 150\,Myr
\citep[a depletion time typical of some longer lived DSFGs at
  high-redshift;][]{swinbank14a}, and 500\,Myr \citep[a DSFG timescale
  which would rely on some sustained gas fueling, which some assert is
  likely the case at the massive end of the galaxy
  `main-sequence;'][]{elbaz11a}.  This figure illustrates that the
assumed timescale for DSFGs and luminous AGN is rather important to
our understanding of cluster assembly.  Over a 2\,Gyr build time, if
DSFGs/AGN are short-lived then the probability of observing $>$5 such
sources in one proto-cluster structure is $<$0.5\%\ (50\,Myr),
6.1\%\ (100\,Myr), 19\%\ (150\,Myr), 77\%\ (500\,Myr).  However,
structures like the COSMOS $z=2.47$ structure and SSA22 contain 12
rare sources {\it each}.

With a short-lived phase, this is virtually
impossible through uncorrelated triggering ($<$1$\times$10$^{-4}$\%),
and still yet unlikely for long duration events ($<$25\%).
If such phenomena are short-lived, then they most certainly are
triggered simultaneously in an event that stretches across very large
volumes.  One can imagine this triggering is brought on by the rapid
collapse of filamentary structure that extends across several tens of
Mpc.

On the other hand, the test above seems to suggest that longer
lifetimes are far more likely (by over a factor of ten) for DSFGs and
luminous AGN in proto-clusters.  Recent simulations work
\citep{narayanan15b} suggest that even somewhat isolated DSFGs could
sustain sufficiently high star-formation rates ($\simgt$500\sfr) for
0.75\,Gyr.  Physically, this sounds plausible particularly in dense
environments, where high star-formation rates may be sustained over
longer periods of time if the galaxies are continually fed fresh
supplies of gas from the surrounding, rich medium.  In the next few
subsections, I explore observations which support both rapid collapse
and heightened gas supply scenarios.

\input{tabco}

\subsection{Molecular Gas Depletion Time}\label{sec:co}

Determining the correct interpretation of the assembly history of
galaxy clusters requires direct constraints of the molecular gas
potential wells in proto-cluster DSFGs.  This gives critical
information on galaxies' current gas supply, and over what time period
such high star-formation rates would be continuously sustainable.  To
reiterate, this is a particularly useful measurement in DSFGs due to
their rarity, as demonstrated in the previous section.

Table~\ref{tab:co} summarizes existing CO observations of proto-cluster
DSFGs from the literature.  Though limited in number and heterogeneous
in transition and depth, these data can begin to discern the
plausibility of short-lived versus long-lived interpretations.
However, as with most previous work on high-$z$ CO observations it is
very important to recognize that the conversion from observed CO line
strength to H$_2$ gas mass is highly uncertain.  It first requires a
conversion from a high-J CO transition to the ground state CO(1-0),
which requires knowledge of the galaxy's mean CO excitation ladder, or
kinetic gas temperature. Second, the conversion from CO(1-0) to M$_{\rm
  H_{2}}$, known as $X_{\rm CO}$ or $\alpha_{\rm CO}$, can also range
by factors of 5--10 depending on gas conditions in the ISM.  For
example, the Milky Way has a gas conversion rate of $\alpha_{\rm
  CO}=$4.5\,M$_\odot$\,(K\,km\,s$^{-1}$\,pc$^2$)$^{-1}$
\citep{bloemen86a,solomon87a} while typical local ULIRGs have
$\alpha_{\rm CO}=$0.8\,M$_\odot$\,(K\,km\,s$^{-1}$\,pc$^2$)$^{-1}$
\citep{downes98a}.  The uncertainties in these two conversions alone
can account for a factor of $\simgt$10 in the predicted gas mass,
which could dramatically affect the interpretation of the depletion
timescale, $\tau_{\rm depl}=M_{\rm H_2}/$SFR.

\begin{figure}
\includegraphics[width=0.99\columnwidth]{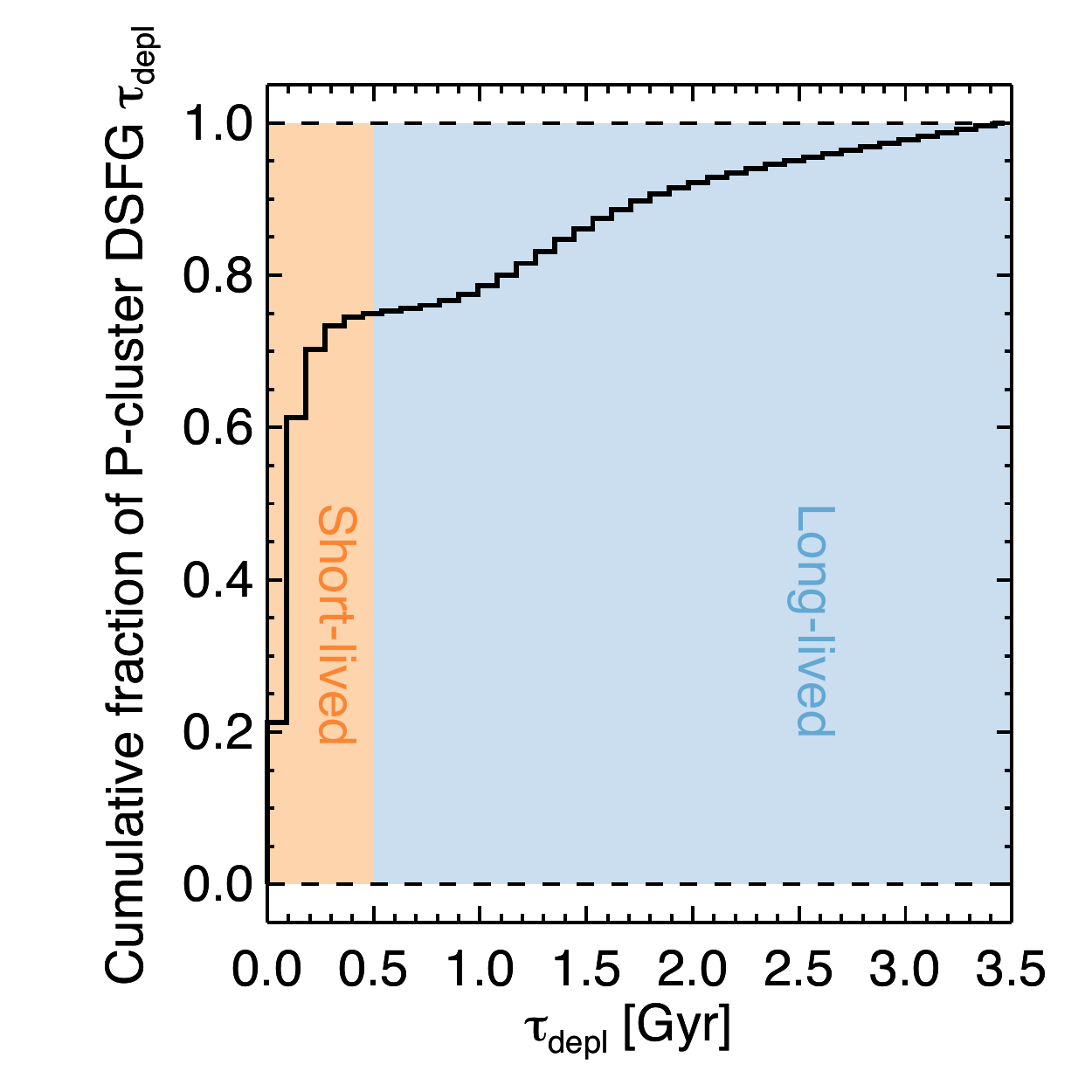}
\caption{The cumulative distribution of gas depletion times for DSFGs
  in proto-clusters as given in Table~\ref{tab:co}.  Each of the eight
  source's depletion times is represented as a Gaussian with
  associated uncertainty.  Here $\tau_{\rm depl}=M_{\rm H_2}/{\rm
    SFR}$, and M$_{\rm H_2}$ is estimated from observations of
  CO. While only eight proto-cluster DSFGs in the literature have CO
  measurements, the majority are consistent with short depletion
  times, $\simlt$200\,Myr (60\%), supporting the idea that
  proto-clusters endure wide-scale star-forming bursts.}
\label{fig:tau}
\end{figure}

For those proto-cluster DSFGs without CO(1-0) observations, a CO gas
excitation ladder, and associated uncertainties, is assumed as given
in \citet{bothwell13a}, the median excitation seen in all observed
DSFGs to-date.  Their figure~3 shows this median DSFG spectral
line-energy distribution.  Each high-J CO line luminosity in
Table~\ref{tab:co} is thus converted to an estimated CO(1-0) line
luminosity via L$_{\rm CO(1-0)}^\prime$/L$_{\rm
  CO(J-[J-1])}^\prime$\,=\,$(S_{\rm CO(1-0)}/S_{\rm
  CO(J-[J-1])})(1/J)^2$.  The uncertainty in the CO Spectral Line
Energy Distribution (SLED) is reflected in the resulting uncertainty
of CO(1-0) line luminosity.  The conversion from $L_{\rm
  CO(1-0)}^\prime$ to M$_{\rm H_2}$ assumes $\alpha_{\rm
  CO}=$\,1.0\,M$_{\rm H_2}$\,(K\,km\,s$^{-1}$\,pc$^2$)$^{-1}$, the
same value adopted in \citet{bothwell13a} and justified generally
through some limited dynamical mass constraints.  The resulting gas
masses M$_{\rm H_2}$ are given in Table~\ref{tab:co}, with some
proto-cluster DSFGs containing multiple components.  In the case where
multiple high-J CO transitions are observed for a single galaxy, a
molecular gas mass is derived for each independently, then averaged.
Depletion times are then calculated by taking the total molecular gas
mass estimated to be present in the system and dividing by the current
star-formation rate, as calculated in \S~\ref{sec:observations}.  The
probability distribution in depletion times is shown in
Figure~\ref{fig:tau}.  Though quite sparse, the majority of sources
(5/7\,$\approx$\,71\%) are estimated to be short-lived, with
$\tau_{\rm depl}\simlt\,150$\,Myr.

\subsection{Evidence supporting rapid bursts in Proto-clusters}

The discussion presented on the measured molecular gas depletion time
of DSFGs in proto-clusters heavily favors a rapid collapse model,
whereby the massive galaxies in clusters are built in short-lived,
extreme episodes that {\it permeate the entire volume of the
  not-yet-virialized proto-cluster.}  The measured gas depletion times
for proto-cluster DSFGs (as presented in Table~\ref{tab:co} and
Figure~\ref{fig:tau}) are the most crucial constraint to this
argument, but it is significantly strengthened by inferred constraints
on the lifetimes of AGN with comparable luminosities to unobscured
quasars \citep{marconi04a}.  The strong evidence for short-lifetimes,
combined with the low probability of observing $N\simgt5$ of these
rare galaxies in one structure, argue for correlated, simultaneous triggering.
Such simultaneous triggering has been directly demonstrated in smaller
isolated cases, as in \citet{ivison13a}, though not on physical scales this large.
If correct, the result is rather extraordinary, as it represents the
only type of direct observation of a temporal `event' on cosmological
scales, spanning a volume $\sim$10$^{4}$\,cMpc$^3$.  In the
next subsection, I briefly explore evidence which supports the
contrary conclusion.

\subsection{Evidence in favor of Gradually-Built Proto-clusters}

Though analysis of literature DSFGs in proto-clusters suggests they are
mostly short-lived, the impact of our high-J CO to gas mass
assumptions should be revisited.  If our assumptions were to be revised
in favor of more `Milky Way' type gas excitation and higher intrinsic
value of $\alpha_{\rm CO}$, the CO(1-0) line luminosities would be a
factor of $\sim$3 higher, and the gas masses a factor of $\sim$20
higher.  The median depletion time of 110\,Myr would instead be
2.2\,Gyr, much more in line with the predicted long-lifetimes of DSFGs
in some cosmological simulations \citep{narayanan15b}.  Reducing the
intrinsic uncertainty in this measurement requires CO(1-0)
measurements of a larger sample of proto-cluster DSFGs with additional
resolved dynamical mass constraints to hone in on the correctly
applicable $\alpha_{\rm CO}$.  Some of these observations are
currently underway at the Jansky Very Large Array.  However, it should
be noted that there is a known upper limit to how long DSFGs can
sustain high-SFRs, given by stellar mass constraints for the
Universe's most massive galaxies.  For example, a galaxy cannot
reasonably maintain a 500\,\sfr\ star-formation rate for longer than
1\,Gyr or so, else the mass of stars produced will exceed
5$\times10^{11}$\,\msun.

Another possible caveat to our rapid collapse argument is the possibly
heightened replenishment of gas reservoirs from the IGM.  It has
recently become clear that galaxies recycle gas through ejective
feedback and outflows, and the eventual reaccretion of material on
$\sim$Gyr timescales \citep{christensen15a}; however, it is unclear
how dense environments at the intersections of filaments in the IGM
might shorten the gas recycling timescale and potential heightened
inflow of pristine material.  If molecular gas is fed onto galaxies
more efficiently in proto-clusters than in the field, particularly on
$\sim$100\,Myr timescales, then the depletion time measurement might
not be an accurate reflection of the lifetime of high-SFR systems.
However, such a dramatically fast ($\simlt$\,100\,Myr) replenishment
of $\sim$10$^{10}$\,\msun\ gas reservoir is unlikely, again due to the
upper limit placed on high-SFR timescale from observed stellar mass
functions.

Finally, as mentioned at the end of \S~\ref{sec:common}, the frequency
of DSFG-rich proto-clusters among the population of all proto-clusters
raises a potential concern.  If the timescale of the DSFG-rich phase
is short-lived and unique, then one may only expect a small subset of
observed $z>2$ proto-clusters to have such DSFG excesses.  To gauge the
plausibility of this argument, we should consider how many member
galaxies we {\it expect} to go through such a phase over the course of
a cluster's lifetime.  In \S~\ref{sec:simultaneous}, this was
approximated as 40$\pm$10 massive $>$10$^{11}$\,\msun\ galaxies.  If
there are 5--10 rare galaxies per proto-cluster, then we may expect
such structures to go through 4--8 ``episodes'' of heightened activity
before virialization at $z<2$.  If these episodes are assumed to all
occur between $2<z<5$ ($\approx$2\,Gyr) then one would expect
$\sim$20--40\%\ of all proto-clusters of that epoch to be DSFG-rich
assuming a 100\,Myr `burst' lifetime.  With a 150\,Myr lifetime, the
fraction shifts to $\sim$30-60\%, and at 50\,Myr only $\sim$10-20\%.
Though these fractions are certainly non-negligible, it is clear that
it would be nearly impossible for {\it all} $z>2$ proto-clusters to be
DSFG-rich {\it if} they are short-lived and therefore our comparison
to the measured cluster mass function at $z\sim0$ might disfavor short
timescales.  It is certainly clear that refining measurements of the
volume density of high-$z$ proto-clusters is needed before ruling out
different histories of their assembly.

\section{Predictions}\label{sec:predictions}

\subsection{Future Observations}

The most important observational characteristic of massive galaxy
clusters is the large area they subtend on the sky, $\sim$half a
degree across.  While some recent works have recognized the importance
of this \citep[e.g.][]{muldrew15a}, the observational community which works on
proto-cluster science has largely overlooked the shear scale of early,
overdense structures.  It is critical to address this if we desire to
move beyond simple proto-cluster discoveries and learn about the
collapse of large scale structure from an observational perspective.

The next generation of wide field (and sufficiently deep) surveys $-$
on order tens of square degrees $-$ will be of great importance to
identifying statistically large samples ($\sim$100) of proto-clusters,
both those with and without rare galaxies.  The most efficient means
of confirming high-redshift overdensities like these will be through
direct far-infrared/millimeter molecular line detection, which may
only be efficient on large scales with the next generation of submm
single-dish multi-pixel spectrometers.  The relative fraction of such
structures with rare galaxies will, in turn, allow the more detailed
look at all clusters' temporal evolution.

On slightly smaller angular scales, recent work from
\citet{clements14a}, \citet{planck-collaboration15a} and
\citet{flores-cacho16a} search out proto-clusters rich in dusty
star-formation by leveraging the poor spatial resolution of the
\planck\ satellite, which covers the entire sky.  Following-up
\planck's $\sim$5$\arcmin$ point sources with the higher resolution
          {\it Herschel Space Observatory} is hoped to be an efficient
          way of identifying early clusters in formation.  While none
          have yet been spectroscopically confirmed, over 200
          candidate high-redshift clusters have been identified with
          an excess of dusty starbursts peaking at $\simgt$350\um.
          The technique is certainly promising though will be quite
          incomplete in the type of structure discussed in this paper,
          as many dusty starbursts would need to fall in one
          \planck\ beam, much smaller than the previously discussed
          half-degree scale.

In terms of characterizing known structures more fully, narrow-band
imaging should provide the most complete mapping of filamentary
structures on the largest scales.  This is the case in the SSA22
$z=3.09$ structure, as well as some structures not observed in the
submillimeter \citep[e.g. the Bo\"otes $z=3.78$ structure;][]{lee14a}
but has not been pursued over sufficiently large angular scales for
most proto-clusters.  Similarly, wide-field IFU spectroscopic follow-up
will be quite valuable, from facilities like the VIRUS instrument on
the Hobby-Eberly Telescope.

It is clear that understanding galaxies' gas supply is an essential
element in discerning proto-clusters' assembly history, and in the age
of ALMA and the Jansky VLA, is not limited to the most luminous, rare
galaxies.  Scaling of long-wavelength dust continuum to an ISM mass
has shown to be a useful proxy \citep{scoville14a,scoville15a} to
galaxies' star-forming molecular gas masses.  Thus fairly inexpensive
observational campaigns to constrain the gas content of
proto-clusters' normal galaxy members might provide more important
clues as to how environment influences galaxies' evolution.

%SDSS quasar search again?

\subsection{Simulations}

Simulations of large-scale structure collapse on cosmological scales
plays a crucial role in our current picture of galaxy cluster
formation, linking the huge gap between observations of nearby
virialized clusters and the imprint of density perturbations on the
Cosmic Microwave Background.  Large-box $>$100\,Mpc simulations are
certainly needed to analyze $>$10$^{15}$\,\msun\ halos. Their enormous
volumes limit the incorporation of baryonic physics and force the
implementation of ultraluminous starbursts, or luminous AGN, to be
somewhat crude.  Yet, there are some basic measurements which could be
extracted from the current generation of simulations that would shed
ample light on the proposed assembly history of massive galaxy
clusters.
In dark-matter only simulations, the most direct probe of cluster
assembly is the merging of dark matter halos with time.  These merger
trees, mapped with spatial distribution, could directly trace whether
or not growth of halos, and thus the galaxies living in them, is
episodic or steady.

Beyond the measurement of stochasticity in assembly, simulations will
be needed to more accurately constrain dark matter halo masses from
observations.  While it is clear that linear bias assumptions break
down under certain pretexts, it is not entirely appropriate to use
normal abundance matching techniques which are more ideally suited for
isolated halos.  An in-depth look at halo mass distributions in
proto-clusters before virialization might provide crucial insight that
bridges our theoretical understanding to observational constraints.

\section{Conclusions}\label{sec:conclusions}

This paper has employed literature datasets to demonstrate that
several high-redshift proto-cluster environments are rich with rare
galaxies: both dusty star-forming galaxies and ultraluminous AGN.
These proto-clusters subtend 10$\arcmin$ to a half degree in the sky
because they have not yet relaxed into virialized galaxy clusters.  By
virtue of their large occupied volumes at $z\simgt2$ (factors of a few
hundred larger than at $z\sim0$), it is very difficult to detect their
significance via an overdensity of `normal' galaxies on appropriately
large scales, which are only slightly more dense than the field.
Instead, an unexpected excess of rare galaxies ($\simgt$5 per
$\sim$10$^{4}$\,cMpc$^3$ volume) can demonstrate a more compelling
argument for a large-scale proto-cluster in formation.

Five bona-fide DSFG-rich proto-clusters have been identified to-date
within $1.99<z<3.09$.  Estimates to their volume density$-$constrained
by deep spectroscopic and submm datasets$-$is
$\sim$5$\times$10$^{-8}$\,cMpc$^{-3}$, similar to the density of
observed $>$10$^{15}$\,\msun\ clusters at $z<0.2$.  Some simulations
work expect the volume density of DSFG-rich structures to be a factor
of $\sim$5 less than observed.

The rarity of DSFGs and luminous AGN relates to their intrinsically
short duty cycle.  If this population is predominantly short-lived,
then it can be used as a constraint on the assembly history of galaxy
clusters in the time before virialization.  For example, the
probability of observing 10 or more 100\,Myr-duration rare galaxies in
one structure is $<$0.01\%.  This suggests the phenomenon is
exceedingly rare, and yet there are several multiple DSFG-rich
proto-clusters in only a few square degrees of data.  The existence of
these structures provides direct observational evidence that
proto-clusters assemble in short-lived, stochastic bursts that likely
correspond to the collapse of large-scale filaments on 10\,Mpc scales.
In this sense, such episodes represent ``events'' observed on the
largest scales seen since the imprint of recombination from the CMB.

An alternate view may be that the gas potential wells of DSFGs in
proto-clusters are much deeper, fueled by an excess of gas in the
surrounding IGM.  This point of view would argue for more long-lived
DSFGs.  If this is the case, then it {\it is} more likely that DSFGs
in proto-clusters are triggered at somewhat arbitrary times determined
only by their local $<$1\,Mpc surrounds.  As a result, it is also
likely that nearly every observed proto-cluster is DSFG-rich.  The
evidence that supports this claim is our estimate of the volume
density to DSFG-rich proto-clusters and its agreement with the cluster
mass function.  If such DSFGs are short-lived, then at most $\sim$half
of high-$z$ proto-clusters should exhibit an enhanced DSFG-rich phase.

While different threads of evidence support both possible explanations
$-$ short-lived, bursting proto-clusters or gas-enhanced
proto-clusters $-$ measurements of gas depletion times for DSFGs
sitting in these structures suggests they are indeed short-lived.
Therefore the former evolutionary scenario is favored, where DSFG-rich
structures represent a short-lived phase of rapid growth across
incredibly large filaments in the IGM.  More observations of such
structures are needed to constrain the overall population of high-$z$
overdensities, the diversity of their star-formation histories, and to
characterize the galaxies within such structures to learn how galaxy
growth is governed by environment.

\acknowledgements

CMC thanks the anonymous referee for helpful comments and the
University of Texas at Austin, College of Natural Science for support.
CMC also thanks many colleagues for interesting conversations which led
up to this paper, including Chao-Ling Hung, Yi-Kuan Chiang, Asantha
Cooray, Joel Primack, Amy Barger, Len Cowie, Cedric Lacey, William
Cowley, James Bullock, Shea Garrison-Kimmel, Chuck
Steidel, and Nick Scoville.  CMC would also like to acknowledge
financial support and fruitful discussions from the Munich Institute
for Astro- and particle Physics (MIAPP) summer programme on the ``Star
Formation History of the Universe,'' held in Munich in August 2015.

\bibliography{caitlin-bibdesk}

\end{document}

%% file: tabhdf.tex
\begin{table*}
\centering
\caption{FIR Photometric Characteristics of DSFGs in the HDF $z=1.99$ structure}
\begin{tabular}{l@{ }c@{ }c@{ }c@{ }c@{ }c@{ }c@{ }c@{ }c@{ }c@{ }c}
\hline\hline
{\sc Name} & $z$ &  {\sc Alt} & $S_{\rm 250}$ & $S_{\rm 350}$ & $S_{\rm 500}$ & $S_{\rm 850}$ & $S_{\rm 1.4}$ & $L_{\rm IR}$ & SFR & {\sc Ref.} \\
 & &  {\sc Name$^{\diamondsuit}$} & [mJy] & [mJy] & [mJy] & [mJy] & [\uJy] & [$\lsun$] & [\sfr] & \\
\hline
DSFG\,J123600.13$+$621047.2 & 1.994 & SMG-93 & $-$ & 12.9$\pm$4.9 & 13.1$\pm$4.5 & 7.9$\pm$2.4$*$ & 128.5$\pm$8.1 & (1.1$^{+1.0}_{-0.5}$)$\times$10$^{12}$ & 100$^{+170}_{-90}$ & 1 \\
DSFG\,J123618.32$+$621550.5 & 1.994 & SMG132 & 22.9$\pm$4.5 & 30.0$\pm$5.3 & 24.8$\pm$5.4 & 7.3$\pm$1.1 & 172.0$\pm$8.4 & (3.5$^{+1.2}_{-0.9}$)$\times$10$^{12}$ & 330$^{+210}_{-150}$ & 1,4 \\
DSFG\,J123621.25$+$621708.3 & 1.988 & $\dagger$SMG140e+w & 25.1$\pm$4.5 & 19.8$\pm$4.9 & 7.5$\pm$4.8 & 7.8$\pm$1.9 & 169.4$\pm$8.8 & (3.3$^{+1.8}_{-1.2}$)$\times$10$^{12}$ & 310$^{+310}_{-210}$ & 1 \\
DSFG\,J123635.57$+$621424.0 & 2.001 & SMG172 & 21.1$\pm$4.5 & 11.0$\pm$5.0 & $-$ & 5.5$\pm$1.4 & 77.0$\pm$7.8 & (2.8$^{+1.8}_{-1.0}$)$\times$10$^{12}$ & 260$^{+310}_{-170}$ & 1 \\
DSFG\,J123711.99$+$621325.6 & 1.992 & SMG255 & 14.9$\pm$4.5 & 13.1$\pm$5.0 & $-$ & 4.2$\pm$1.4 & 50.2$\pm$8.1 & (2.0$^{+1.8}_{-1.0}$)$\times$10$^{12}$ & 190$^{+310}_{-170}$ & 1,\,2,\,4 \\
DSFG\,J123711.32$+$621330.9 & 1.993 & SFRG254 & 38.0$\pm$4.5 & 34.7$\pm$5.0 & 25.0$\pm$5.0 & $-$ & 79.6$\pm$17.2 & (5.7$^{+3.8}_{-2.3}$)$\times$10$^{12}$ & 540$^{+650}_{-400}$ & 1,\,2,\,4 \\
\hline
RAD\,J123632.53$+$620759.8 & 1.993 & SMG169 & $-$ & $-$ & $-$ & 5.5$\pm$1.3$*$ & 80.4$\pm$8.6 & (3.9$^{+1.5}_{-0.3}$)$\times$10$^{11}$ & 40$^{+30}_{-5}$ & 1 \\
RAD\,J123617.54$+$621540.7 & 1.993 & $\ddag$SFRG130 & $-$ & $-$ & $-$ & $-$ & 200.0$\pm$12.8 &  $-$ & $<$12 & 1,\,3 \\
RAD\,J123640.73$+$621011.0 & 1.977 & SFRG179 & $-$ & $-$ & $-$ & $-$ & 72.5$\pm$8.3 & $-$ & $<$50 & 1 \\
\hline\hline
\end{tabular}
\label{tab:hdf}
\begin{minipage}{\textwidth}
{\small {\bf Table Notes.}  References noted in the last column are
  1=\citet{chapman09a}, 2=\citet{casey09b}, 3=\citet{casey09a},
  4=\citet{bothwell10a}.

$^{\diamondsuit}$ {\sc Alt Name} is the alternate name used for this source
  throughout the literature, and as stated in \citet{chapman09a}.

$*$ The original 850\um\ flux densities as measured by \scuba\ for
  SMG-93 and SMG169 are inconsistent with more recent
  850\um\ follow-up with \scubaii\ \citep{chen13b}.

$\dagger$ Source SMG140e$+$w is a double radio source within a single
  \scuba\ beam; the second radio source has flux density
  63.4$\pm$10.6\,\uJy.  

$\ddag$ Source SMG130 was originally thought to be a
  submillimeter-faint star-forming radio galaxy (SFRG/OFRG) in
  \citet{chapman04a} but was later revealed through high-resolution
  radio imaging to be a low-luminosity AGN in an evolved galaxy
  \citep{casey09a}. }
\end{minipage}
\end{table*}

%% file: tabssa22.tex
\begin{table*}
\centering
\caption{FIR Characteristics of DSFGs in the SSA22 $z=3.09$ structure}
\begin{tabular}{lccccccccc}
\hline\hline
{\sc Name} & $z$ & $S_{\rm 850}$ & $S_{\rm 1.1mm}$ & $L_{\rm IR}$ & SFR & X-ray  & {\sc Ref.} \\
           &     & [mJy]         & [mJy]           & [$\lsun$]      & [\sfr] & AGN & \\
\hline
DSFG\,J221732.41$+$001743.8 & 3.092   & ...          & 6.4$\pm$0.2 & (1.1$^{+0.9}_{-0.5}$)$\times$10$^{13}$ & 1000$^{+800}_{-500}$ & Y & 3 \\
DSFG\,J221735.15$+$001537.3 & 3.096/8 & 6.3$\pm$1.3  & 2.3$\pm$0.1 & (4.5$^{+3.7}_{-2.0}$)$\times$10$^{12}$ & 420$^{+340}_{-190}$ & N & 1,\,3 \\
DSFG\,J221735.83$+$001559.0 & 3.089   & 4.9$\pm$1.3  & 1.8$\pm$0.1 & (3.5$^{+2.9}_{-1.6}$)$\times$10$^{12}$ & 330$^{+270}_{-150}$ & Y & 1,\,3 \\
DSFG\,J221732.01$+$001655.4 & 3.091   & 3.2$\pm$1.6  & 0.7$\pm$0.1 & (1.8$^{+1.4}_{-0.8}$)$\times$10$^{12}$ & 160$^{+140}_{-70}$ & Y & 2,\,3 \\
DSFG\,J221725.97$+$001238.9 & 3.102   & 17.4$\pm$2.9 & $-$         & (1.4$^{+1.2}_{-0.6}$)$\times$10$^{13}$ & 1400$^{+1100}_{-600}$ & N & 1,\,2 \\
%LAB\,J221724.68$+$001242.0  & 3.06-3.13     & 16.8$\pm$2.9 & $-$          & (1.4$^{+1.1}_{-0.6}$)$\times$10$^{13}$ & 1300$^{+1100}_{-600}$ & N \\ % same as source above
LAB\,J221711.67$+$001644.9  & 3.06--3.13 & 5.2$\pm$1.4  & $-$          & (4.3$^{+3.5}_{-1.9}$)$\times$10$^{12}$ & 400$^{+330}_{-180}$ & N & 2 \\
LAB\,J221802.27$+$002556.9  & 3.06--3.13 & 6.1$\pm$1.4  & $-$          & (5.1$^{+4.1}_{-2.3}$)$\times$10$^{12}$ & 480$^{+390}_{-210}$ & N & 2 \\
LAB\,J221728.90$+$000751.0  & 3.06--3.13 & 11.0$\pm$1.5 & $-$          & (9.1$^{+7.5}_{4.1}$)$\times$10$^{12}$  & 860$^{+700}_{-390}$ & N & 2 \\
$\dagger$DSFG\,J221737.11$+$001712.4 & 3.090   & ...          & 1.1$\pm$0.1 & (1.8$^{+1.5}_{-0.8}$)$\times$10$^{12}$ & 170$^{+140}_{-80}$ & N & 3 \\
$\dagger$DSFG\,J221736.54$+$001622.7 & 3.095   & ...          & 1.0$\pm$0.1 & (1.7$^{+1.4}_{-0.7}$)$\times$10$^{12}$ & 160$^{+130}_{-70}$ & Y & 3 \\
$\dagger$DSFG\,J221737.05$+$001822.4 & 3.086   & ...          & 1.1$\pm$0.1 & (1.8$^{+1.5}_{-0.8}$)$\times$10$^{12}$ & 170$^{+140}_{-80}$ & N & 3 \\
$\dagger$DSFG\,J221736.81$+$001818.1 & 3.085   & ...          & 0.8$\pm$0.2 & (1.3$^{+1.1}_{-0.6}$)$\times$10$^{12}$ & 120$^{+100}_{-60}$ & N & 3 \\
\hline\hline
\end{tabular}
\label{tab:ssa}

\begin{minipage}{\textwidth}
{\small {\bf Table Notes.}  Sources with $\dagger$ preceding the name
  are not included in the calculation of SSA22's rare object
  overdensity, $\delta_{\rm rare}$, as they are much lower
  luminosities than the other DSFGs in the sample, detected over much
  larger areas.  References are 1=\citet{chapman05a},
  2=\citet{geach05a}, 3=\citet{umehata15a}.  850\um\ flux densities
  are taken from \citet{chapman05a} and \citet{geach05a} while 1.1\,mm
  flux densities from ALMA are given in \citet{umehata15a}.  Note that
  850\um\ coverage extends over a much larger area than the `ALMA Deep
  Field' of the SSA22 proto-cluster node but with shallower depth.  The
  redshifts of the three LAEs are not precisely known as they were
  identified through narrow-band imaging and not direct spectroscopic
  observations.  The X-ray AGN column indicates whether or not the
  given DSFG is matched to an X-ray source in \citet{lehmer09a}.
  Total infrared luminosities and star-formation rates are derived by
  assuming a 35\,K dust modified blackbody plus mid-infrared powerlaw.
}
\end{minipage}
\end{table*}

%% file: tabco.tex
\begin{table*}
\caption{DSFGs in proto-clusters with CO measurements}
\begin{tabular}{l@{ }l@{ }c@{ }cccccc}
\hline\hline
 DSFG Name & $z$ & Transition & L$_{\rm CO}^\prime$ & M(H$_{\rm 2}$)$^\diamondsuit$ & SFR & $\tau_{\rm depl}$ & Reference \\
& & & [K\,km\,s$^{-1}$\,pc$^2$] & [\msun] & [\sfr] & [Myr] \\
\hline
\multicolumn{2}{c}{\sc Detections:} & & & & & & & \\

%HDF132
DSFG\,J123618$+$621550 & 1.996  & CO(4-3) & (9.4$\pm$1.4)$\times$10$^{10}$ & (2.6$\pm$0.6)$\times$10$^{11}$ & & &  \citet{bothwell10a} \\
                       & 2.001  & CO(4-3) & (6.5$\pm$0.9)$\times$10$^{10}$ & (1.8$\pm$0.4)$\times$10$^{11}$ & & & \citet{bothwell10a} \\
                       &        &         & {\it total:}             & {\it (4.4$\pm$0.7)$\times$10$^{11}$} & 330$^{+110}_{-80}$ & {\bf 1300$\pm$400} & \\
%hdf254
DSFG\,J123711$+$621331 & 1.988  & CO(4-3) & (1.3$\pm$0.2)$\times$10$^{10}$ & (3.6$\pm$0.8)$\times$10$^{10}$ & & & \citet{casey11a} \\
                       & 1.996  & CO(4-3) & (7.8$\pm$1.1)$\times$10$^{9}$  & (2.2$\pm$0.5)$\times$10$^{10}$ & & & \citet{casey11a} \\
                       & 1.995  & CO(3-2) & (1.5$\pm$0.5)$\times$10$^{10}$ & (3.4$\pm$1.2)$\times$10$^{10}$ & & & \citet{casey11a} \\
                       &        &         & {\it average:}           & {\it (4.9$\pm$0.7)$\times$10$^{10}$} & 540$^{+350}_{-210}$ & {\bf 110$\pm$60} & \\
%hdf255
DSFG\,J123712$+$621322 & 1.996  & CO(4-3) & (6.8$\pm$1.5)$\times$10$^{9}$  & (1.9$\pm$0.5)$\times$10$^{10}$ & & & \citet{casey11a} \\
                       & 1.996  & CO(3-2) & (2.7$\pm$0.9)$\times$10$^{10}$ & (6.0$\pm$2.2)$\times$10$^{10}$ & & & \citet{bothwell13a} \\
                       &        &         & {\it average:}           & {\it (2.1$\pm$0.5)$\times$10$^{10}$} & 190$^{+170}_{-90}$ & {\bf 110$\pm$80} & \\
%smg169
DSFG\,J123632$+$620800 & 1.994  & CO(3-2) & $^{\sharp}$(4.0$\pm$1.1)$\times$10$^{10}$ & (8.9$\pm$2.8)$\times$10$^{10}$ & 36$^{+144}_{-30}$ & $\simlt$9000 & \citet{bothwell13a} \\
%\hline 
%spiderweb galaxy
DSFG\,J114048$-$262908 & 2.163  & CO(1-0) & (6.5$\pm$0.6)$\times$10$^{10}$ & (6.5$\pm$0.6)$\times$10$^{10}$ & & & \citet{emonts13a} \\
                       & 2.150  & CO(1-0) & (6.9$\pm$2.3)$\times$10$^{9}$  & (6.9$\pm$2.3)$\times$10$^{9}$  & & & \citet{emonts13a} \\
                       &        &         &                                &                                & $^{\ddag}$740$\pm$80 & {\bf 97$\pm$13} & \citet{seymour12a} \\
%hae229
DSFG\,J114046$-$262913 & 2.147 & CO(1-0) & (3.3$\pm$0.2)$\times$10$^{10}$ & (3.3$\pm$0.2)$\times$10$^{10}$ & $\dagger$480$^{+150}_{-110}$ & {\bf 68$\pm$5} &  \citet{emonts13a} \\
%\hline
DSFG\,J221735$+$001537 & 3.096  & CO(3-2) & (3.8$\pm$1.0)$\times$10$^{10}$ & (8.5$\pm$2.5)$\times$10$^{10}$ & $^{\heartsuit}$1100$^{+300}_{-200}$ & {\bf 80$\pm$30} & \citet{greve05a} \\
DSFG\,J221726$+$001239 & 3.102  & CO(4-3) & (6.7$\pm$2.1)$\times$10$^{10}$ & (1.9$\pm$0.7)$\times$10$^{11}$ & 1400$^{+1100}_{-600}$ & {\bf 140$\pm$90} & \citet{chapman04c} \\
DSFG\,J221732$+$001744 & 3.092  & CO(3-2) & ... & ... & $^{\flat}$1180$^{+890}_{-230}$ & ... & (Yun \etal, in prep) \\
%\hline
\multicolumn{2}{c}{\sc Non-detections:} & & & & & & & \\
%smg140e+w
DSFG\,J123621$+$621708 & 1.973--2.008 & CO(4-3) & $<$5.2$\times$10$^{9}$  & $<$1.5$\times$10$^{10}$ & 310$^{+170}_{-110}$ & $<$50 & \citet{bothwell13a} \\
%smg93
DSFG\,J123600$+$621047$^{\natural}$ & 1.971--2.017 & CO(3-2) & $<$2.9$\times$10$^{10}$ & $<$6.5$\times$10$^{10}$ &          &      & \citet{greve05a} \\
                             &              & CO(3-2) & $<$1.6$\times$10$^{10}$ & $<$3.6$\times$10$^{10}$ &          &      & \citet{bothwell13a} \\
                             &              &         &                         &                         & 110$^{+100}_{-50}$ & $<$300 &  \\
\hline\hline
\end{tabular}
\label{tab:co}
{\small {\bf Table Notes.}

$^{\diamondsuit}$ Gas masses estimated assuming a fixed $\alpha_{\rm CO}$ gas
  conversion factor of $\alpha=1.0$ \citep[as in][]{bothwell13a}.

$^{\dagger}$ tentative detection of CO.

$^{\ddag}$SFR for the Spiderweb galaxy is calculated from the starburst component of the FIR SED as presented in \citet{seymour12a}.

$^\sharp$ SFR for HAE source at $z=2.147$ is taken from H$\alpha$
  measurements from \citet{kurk04a} as 23$\pm$1; however, we refit the
  SFR given the FIR photometry measured in \citet{dannerbauer14a}
  (their table 4).

$^{\heartsuit}$SFR for the SSA22 galaxy calculated from 850um flux density and radio flux density using an SED with temperature 35K from \citet{chapman05a}, also accounting for a deboosting factor $\sim$1.5, consistent with more recent submm datasets.

$^{\flat}$ SFR calculated as in \citet{umehata15a}.

$^{\natural}$ SMG-93, a.k.a. SMM\,J123600$+$621047 is mistakenly labeled as
  SMM\,J123600$+$620253 in \citet{bothwell13a}, but all of the
  physical characteristics listed in \citeauthor{bothwell13a} are
  indeed for SMG-93.

}
\vspace{5mm}
\end{table*}